\documentclass[aps,prb,superscriptaddress,twocolumn,longbibliography]{revtex4-2}
\usepackage{graphicx,xcolor,amsmath,amssymb,physics}
\usepackage{dsfont}
\usepackage[colorlinks,citecolor=blue,urlcolor=blue]{hyperref}

\begin{document}


\title{DC conductivity of tilted Dirac Fermions across the Lifshitz Transition:\\ short- versus long-range impurities}

\author{Mohammad H. Pakzamir}
\affiliation{Department of Physics, Institute for Advanced Studies in Basic Sciences (IASBS), Zanjan 45137-66731, Iran}

\author{Zahra Faraei}\email{zahra.faraei@gmail.com }
\affiliation{Department of Physics, Institute for Advanced Studies in Basic Sciences (IASBS), Zanjan 45137-66731, Iran}

\author{Ali G. Moghaddam}
\address{Department of Applied Physics, Aalto University, 02150 Espoo, Finland}
\address{Computational Physics Laboratory, Physics Unit, Faculty of Engineering and
Natural Sciences, Tampere University, P.O. Box 692, FI-33014 Tampere, Finland}

\date{\today}

\begin{abstract}
We theoretically investigate the DC conductivity of two-dimensional tilted Dirac systems subject to short- and long-range impurity scattering. Using the Kubo formalism, we systematically study transport across the subcritical (Type I), critical, and overcritical (Type II) tilt regimes. In the subcritical phase, short-range impurities yield a frequency-independent conductivity that decreases monotonically with tilt. Conversely, long-range Coulomb scattering results in a strongly energy-dependent conductivity governed by a tilt-independent scattering rate. At the Lifshitz transition ($t=1$), the transport signatures of these impurities diverge fundamentally: the van Hove singularity in the density of states induces a localized conductivity dip for short-range disorder, but a pronounced macroscopic peak for Coulomb impurities. In the overcritical regime, an ultraviolet momentum cutoff is required to regularize the open Fermi surface, leading to distinct behaviors for each impurity type. Notably, the conductivity perpendicular to the tilt direction ($\sigma_{xx}$) exhibits a cutoff-dependent, non-monotonic peak near $t=\sqrt{2}$ for short-range defects, while it decays monotonically with increasing tilt for long-range scattering. For both potentials, the conductivity along the tilt axis ($\sigma_{yy}$) increases without bound, revealing extreme transport anisotropy. For long-range impurities, the energy dependence of the conductivity becomes nearly quadratic and linear for Type I and II, respectively. Furthermore, vertex corrections vanish identically at the Lifshitz transition for both impurity types. Finally, we provide a unified geometric framework for these phenomena, establishing the tilt parameter as a powerful knob for engineering macroscopic transport in Dirac materials.
\end{abstract}

\maketitle

\section{Introduction}

The discovery of gapless topological states—ranging from two-dimensional (2D) graphene to three-dimensional (3D) Dirac and Weyl semimetals—has established a profound paradigm for exploring relativistic quantum phenomena in solid-state systems \cite{novoselov2005two, castro2009electronic, armitage2018weyl, yan2017topological,Ding_RMP}. In these materials, low-energy excitations are accurately described by the massless Dirac equation, which underpins their remarkable transport properties. A hallmark of such systems is the existence of a finite minimum conductivity that remains robust even in the limit of vanishing carrier concentration \cite{ando2002dynamical, nomura2007quantum, DasSarma2011}. Beyond this, transport in these systems exhibits a suite of exotic signatures, including the chiral anomaly, negative longitudinal magnetoresistance, the planar Hall effect, and unique features arising from topological Fermi arc surface states \cite{Turner2011,Burkov_2015,Burkov_2015_NMR,hasan2017discovery,Tewari2017,Burkov_PHE,Nissinen2020,ong2021experimental,Rostamzadeh2022}.

However, this conventional isotropic picture changes significantly when the Dirac cone is tilted in momentum space. Tilted Dirac cones arise naturally in layered organic conductors \cite{katayama2006pressure, goerbig2008tilted} and in Type II Weyl semimetals such as $\mathrm{WTe}_2$ and $\mathrm{MoTe}_2$ \cite{soluyanov2015type, volovik2017lifshitz, wang2016mote2}. The tilt, quantified by the dimensionless parameter $t$, breaks the system’s continuous Lorentz invariance \cite{xu2017discovery,Seradjeh2022}, and leads to interesting transport and dynamical signatures \cite{Bardarson2017,Bergholtz2017,Kundu_2020,Agarwal2019,Ogata2022,Adagideli2019,Mohajerani_2021,NandaarXiv}. Depending on its magnitude, the system falls into two distinct topological phases: the Type I subcritical regime ($t<1$), where the Fermi surface remains a single point or an ellipse, and the Type II overcritical regime ($t>1$), characterized by overtilted cones in which electron and hole pockets coexist at the Fermi level \cite{trescher2015quantum, carbotte2016dirac}. The boundary between these phases is the Lifshitz transition at the critical point $t=1$, where the Fermi-surface topology undergoes a singular transformation \cite{volovik2017lifshitz, mukherjee2019doping}. Interestingly, the connection between the effective spacetime geometry and the tilt parameter has sparked strong interest in exploring concepts such as emergent gravity and black-hole analogs in Dirac/Weyl systems \cite{Volovik:2016kid, volovik2003universe,kedem2020black,Davis2022,guan2017artificial,huang2018black,Akbari2019,Faraei2019,Ojanen2019,Morice2021,moghaddam2021engineering,Sabsovich2021,Wezel2022,konye2023lensing,chowdhury2026dirac}.

Understanding the DC electronic conductivity in these tilted systems requires a careful analysis of the underlying impurity scattering mechanisms. Transport is fundamentally governed by the interplay between the modified density of states (DOS) and the specific nature of the disorder potential \cite{ostrovsky2006electron, sharma2017weyl}. Two common phenomenological models for disorder are short-range potentials (for example, point-like neutral defects) and long-range Coulomb potentials (for example, screened ionized impurities) \cite{DasSarma2011, adam2007two, Bruus2004}. Recently, numerical investigations exploring finite-size scaling and spectral statistics have highlighted the rich localization physics induced by the tilt parameter under simple scalar disorder \cite{NandaarXiv}. However, an analytical computation of the bulk macroscopic DC conductivity in the thermodynamic limit, incorporating physically realistic models for both short-range (point-like) and long-range (Coulomb) scattering potentials across the full range of tilt regimes, remains a significant theoretical challenge. This is particularly important in the overcritical regime, where the open Fermi surface requires careful theoretical regularization.

In this paper, we present a comprehensive and systematic study of the DC conductivity in a two-dimensional (2D) tilted Dirac system. Employing the Kubo formalism within the first-order Born approximation, we evaluate the conductivity tensor for both short-range and long-range Coulomb impurity potentials. By exploring the full parameter space, our primary focus is to uncover the transport signatures as a function of Fermi energy ($\varepsilon_F$) and the continuous evolution of the tilt parameter ($t$) across the subcritical, critical, and overcritical regimes.

The key analytical and numerical findings of this investigation reveal profound transport signatures characterizing the interplay between disorder and topology:

In the \emph{subcritical regime} ($t<1$), for short-range impurities, the longitudinal conductivity maintains an energy-independent plateau that decays monotonically as $t \to 1$. In stark contrast, long-range Coulomb scattering yields a mathematically exact cancellation of the tilt dependence in the self-energy, leading to a scattering rate proportional to $1/|\omega|$. This results in a conductivity that exhibits a smooth, nearly quadratic dependence on the Fermi energy, which heavily suppresses the transport near the Dirac point. This regime for short-range impurities has been previously explored in other works but assuming a constant scattering rate \cite{Adagideli2019}.

At the \emph{critical point} ($t=1$), corresponding to the Lifshitz transition, the nature of transport critically distinguishes the two scattering potentials. For short-range impurities, the van Hove singularity in the density of states drives a severe divergence in the scattering self-energy, leading to a sharply localized suppression (dip) in the macroscopic DC conductivity. In stark contrast, for long-range Coulomb scattering, the momentum dependence of the matrix elements effectively balances the geometric divergence, leading to a robust enhancement (peak) of the conductivity as the system approaches the Lifshitz transition. Furthermore, we analytically demonstrate that vertex corrections (accounting for multiple-scattering events) vanish identically exactly at $t=1$ for both potentials.

In the \emph{overcritical regime} ($t>1$), the emergence of an open hyperbolic Fermi surface necessitates the introduction of an ultraviolet momentum cutoff $\Lambda$ to regularize the transport integrals, which subsequently leads to a fundamental distinction between the two scattering mechanisms. For short-range impurities, the scattering rates and 
conductivity exhibit an explicit dependence on the cutoff $\Lambda$. Notably, the conductivity in perpendicular direction to the tilt, denoted by $\sigma_{xx}$, displays an anomalous non-monotonic bechavior: it reaches a distinct maximum near $t\approx\sqrt{2}$ before undergoing an asymptotic $1/t$ decay at large tilts due to severe phase-space constriction. But the conductivity component along the tilt direction, $\sigma_{yy}$, increases without bound. This reveals an extreme transport anisotropy characterized by $\sigma_{xx}\gg \sigma_{yy}$, serving as a defining hallmark of Type II Dirac fermions.
Conversely, for long-range impurities, the conductivity decays from its maximum at $t=1$ as the tilt increases. We note that although the scattering rates remain intrinsically ultraviolet-finite in this case, the open hyperbolic Fermi lines extended to infinity still requires a cutoff. 

These results establish the tilt parameter as a versatile tuning knob for engineering topological and transport phase transitions in Dirac semimetals. The structure of this paper is organized as follows. In Sec. \ref{sec:theory}, we introduce the theoretical formalism, the tilted Hamiltonian, and the analytical derivation of vertex corrections. Sec. \ref{sec:shortrange} discusses the self-energies and the global conductivity behavior for the short-range impurity potential across all three tilt regimes. In Sec. \ref{sec:longrange}, we extend this analysis to the long-range Coulomb potential. Finally, in Sec. \ref{sec:geometric}, we provide a geometric reinterpretation of our results by reformulating the tilt parameter as an effective metric curvature in momentum space, which unifies the analytical findings. Finally, we provide a discussion over all the findings in Sec. \ref{sec:discussion} before the summary. 

\section{Theoretical model and formulation}\label{sec:theory}

\subsection{Low-energy Hamiltonian}
We consider a two-dimensional tilted Dirac system described, in units with $\hbar=v_F=1$, by the continuum Hamiltonian
\begin{align}
H(\mathbf{q}) = t q_y \mathbb{I}_{4}
-\tau_z\otimes\left(q_y\sigma_x+q_x\sigma_y\right),
\label{eq:H_tilted_Dirac}
\end{align}
where $\mathbf{q}=(q_x,q_y)$ is the momentum measured from the Dirac point. The matrices
$\tau_x,\tau_y,\tau_z$ and
$\sigma_x,\sigma_y,\sigma_z$ are two independent sets of Pauli matrices acting on different internal degrees of freedom, such as valley, spin, or sublattice. The $4\times4$ identity matrix is denoted by $\mathbb{I}_{4}$. The dimensionless parameter $t$ controls the tilt of the Dirac cone along the $q_y$ direction. For $t=0$, Eq.~\eqref{eq:H_tilted_Dirac} reduces to the untilted Dirac Hamiltonian, while the Lifshitz transition occurs at $|t|=1$; for $|t|>1$, the Dirac cone becomes overtilted.

Diagonalization of Eq.~\eqref{eq:H_tilted_Dirac} gives the two-band spectrum
\begin{align}
\varepsilon_s(\mathbf{q}) = t q_y + s q,
\qquad
q\equiv\sqrt{q_x^2+q_y^2},
\qquad
s=\pm,
\label{eq:dispersion}
\end{align}
where $s=+1$ and $s=-1$ label the conduction and valence bands, respectively. Each band is twofold degenerate due to the $\tau$ degree of freedom.

A convenient normalized set of eigenstates is
\begin{align} 
\Psi_{+, -}(\mathbf{q})
&=
\frac{1}{\sqrt{2}}
\begin{pmatrix}
\chi_{\mathbf{q}} \\ 1 \\ 0 \\ 0
\end{pmatrix},
&
\Psi_{+, +}(\mathbf{q})
&=
\frac{1}{\sqrt{2}}
\begin{pmatrix}
- \chi_{\mathbf{q}} \\ 1 \\ 0 \\ 0
\end{pmatrix},
\nonumber \\
\Psi_{-, +}(\mathbf{q})
&=
\frac{1}{\sqrt{2}}
\begin{pmatrix}
0 \\ 0 \\ \chi_{\mathbf{q}} \\ 1
\end{pmatrix},
&
\Psi_{-, -}(\mathbf{q})
&=
\frac{1}{\sqrt{2}}
\begin{pmatrix}
0 \\ 0 \\ - \chi_{\mathbf{q}} \\ 1
\end{pmatrix}, 
\label{eq:eigenstates}
\end{align}
where the first subscript denotes the eigenvalue of $\tau_z$, and the second subscript denotes the band index $s=\pm$. The phase factor $\chi_{\mathbf{q}}$ is defined as $\chi_{\mathbf{q}}
=(q_y-iq_x)/q$ which satisfies $|\chi_{\mathbf{q}}|=1$. The four states in Eq.~\eqref{eq:eigenstates} form a complete orthonormal basis for $\mathbf{q}\neq 0$.

\begin{figure}[t]
	\includegraphics[width=0.95\linewidth]{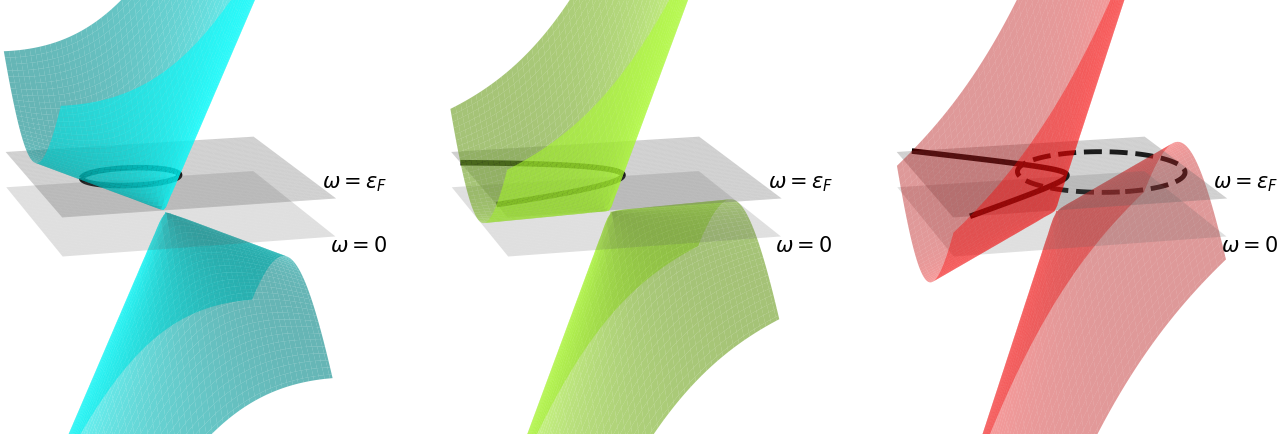}
	\caption{Schematic evolution of the Fermi surface topology in a tilted Dirac semimetal at a constant energy $\omega > 0$. The contours represent the subcritical closed ellipse ($t<1$, blue), the critical Lifshitz transition point ($t=1$, green), and the overcritical open hyperbolas ($t>1$, red). . The dashed lines illustrate the transverse momentum cutoff  $\Lambda$ required to regularize transport in the overcritical regime.}
	\label{fig:Fermi_Surface_Schematic}
\end{figure}

The variation of the tilt parameter $t$ fundamentally alters the geometric topology of the Fermi surface, driving a Lifshitz transition at $t=1$. To provide a clear physical intuition for the scattering phase space, we schematically illustrate the constant energy contours in Fig.~\ref{fig:Fermi_Surface_Schematic}. For $t<1$ (subcritical), the Fermi surface is a closed ellipse, ensuring bounded momentum integration. At the critical point $t=1$, the contour transforms into a parabolic-like open shape (approaching parallel lines at low energies). For $t>1$ (overcritical), the surface splits into two disconnected, open hyperbolic branches. The unbounded nature of these hyperbolas necessitates the introduction of an ultraviolet momentum cutoff $\Lambda$ to regularize the transport integrals, mirroring the intrinsic bandwidth limits in real materials. 

\subsection{Conductivity and the Kubo formula}
The conductivity tensor within linear response theory is given by the Kubo formula
\cite{burkov2011topological}:
\begin{align}\label{eq:Kubo}
	\sigma _{\alpha \beta}=&\frac{e^2}{\pi}	\int_{-\infty}^{+\infty}d\varepsilon \left( -
  \partial_\varepsilon f \right) 
  \sum_{\lambda\lambda'} \int  \frac{d^2q }{(2\pi)^2} \: 
  \bra{\lambda\mathbf{q}}\hat{v}_{\alpha}{\ket{\lambda^{\prime}\mathbf{q}}}
  \nonumber\\ 
  &\qquad \times  {\bra{\lambda^{\prime}\mathbf{q}}}\hat{v}_{\beta}{\ket{\lambda\mathbf{q}}}\,\mathrm{Im}\, G_{{\lambda}}^R(\mathbf{q}, \varepsilon)\,\mathrm{Im}\, G_{{\lambda^{\prime}}}^R(\mathbf{q}, \varepsilon),
\end{align}
where $\lambda$ and $\lambda'$ denote band indices (conduction or valence), $\hat{v}_{\alpha}={\partial{H}({\mathbf{q})}}/{\partial q_{\alpha}}$ (setting $\hbar=1$) and $f(\varepsilon)$ is the Fermi-Dirac distribution function. In the zero temperature limit, the energy integral simplifies as, $\partial_\varepsilon f  \rightarrow \delta(\varepsilon - \varepsilon_F)$ where $\varepsilon_F$ is the Fermi energy. The velocity operators are evaluated in the basis of \eqref{eq:eigenstates}. 

The retarded Green's function for band $\lambda$ is written as
\begin{align}
G^R_{\lambda}(\mathbf{q},\omega)
=
\frac{1}
{\omega+i0^+-\varepsilon_{\lambda}(\mathbf{q})
-\Sigma^R_{\lambda}(\mathbf{q},\omega)} ,
\end{align}
where $\varepsilon_{\lambda}(\mathbf{q})$ denotes the eigenvalue of the unperturbed Hamiltonian $H ({\mathbf q}$), and
$\Sigma^R_{\lambda}(\mathbf{q},\omega)$ is the retarded self-energy in the band basis.

Within the Born approximation, the disorder-induced self-energy is obtained from the impurity scattering potential as
\begin{align}
\hat{\Sigma}^R(\mathbf{q},\omega)
=
n_{\mathrm{\rm imp}}
\int \frac{d^2 q'}{(2\pi)^2}
\left|V(\mathbf{q}-\mathbf{q}')\right|^2
\hat{G}_0^R(\mathbf{q}',\omega),
\end{align}
where $n_{\mathrm{\rm imp}}$ is the impurity concentration, $V(\mathbf{q}-\mathbf{q}')$ is the Fourier component of the impurity potential, and
$\hat{G}_0^R$ is the bare retarded Green's function. In the band representation, the bare Green's function is diagonal and has the form
\begin{align}
G^R_{0,\lambda}(\mathbf{q},\omega)
=
\frac{1}
{\omega+i0^+-\varepsilon_{\lambda}(\mathbf{q})}.
\end{align}
The imaginary part of the retarded self-energy $\Gamma_{\lambda}(\mathbf{q},\omega)
= \operatorname{Im}
\Sigma^R_{\lambda}(\mathbf{q},\omega)$ determines the quasiparticle broadening or equivalently scattering rate. For the tilted Dirac model considered here, the conduction and valence bands are each twofold degenerate. Consequently, in the diagonal band basis there are only two independent quasiparticle broadening functions $\Gamma_{\mathrm{I}}$ and $\Gamma_{\mathrm{II}}$ associated with the valence and conduction bands, respectively.

Substituting the Green's functions into the Kubo formula and performing the frequency integration at zero temperature, we obtain the longitudinal components of the dc conductivity as
\begin{align}\label{eq:conductivity_integrand}
\sigma_{\alpha\alpha}
=
\frac{e^2}{\pi}
\int\frac{d^2 q}{(2\pi)^2}
\sum_{i} C_{\alpha\alpha}^{(i)}(\mathbf q,\omega=\varepsilon_F),
\end{align}
where the integrands $C_{\alpha\alpha}^{(i)}(\mathbf q,\omega)$ are determined by the velocity matrix elements and the band-resolved self-energies. For the $xx$ component, they take the explicit form
\begin{align}
C_{xx}^{(1)}(\mathbf q,\omega)
&=
\zeta_y\,
\frac{4\,\Gamma_{\mathrm{I}}\Gamma_{\mathrm{II}}}
{\left(\Gamma_{\mathrm{II}}^{2}+\xi_{-}^{2}\right)
\left(\Gamma_{\mathrm{I}}^{2}+\xi_{+}^{2}\right)},
\notag \\
C_{xx}^{(2)}(\mathbf q,\omega)
&=
\zeta_x\,
\frac{2\,\Gamma_{\mathrm{II}}^{2}}
{\left(\Gamma_{\mathrm{II}}^{2}+\xi_{-}^{2}\right)^{2}},
\notag \\
C_{xx}^{(3)}(\mathbf q,\omega)
&=
\zeta_x\,
\frac{2\,\Gamma_{\mathrm{I}}^{2}}
{\left(\Gamma_{\mathrm{I}}^{2}+\xi_{+}^{2}\right)^{2}}, \label{eq:Cxx}
\end{align}
where
$\xi_{\pm}
= 
\omega-tq_y\mp q$ and $\zeta_i
=
\left(\frac{q_i}{q}\right)^{2}$ for $i=x,y$. Analogous expressions hold for the $yy$ component, obtained by replacing the prefactor $\zeta_y$ in $C_{xx}^{(1)}$ with $\zeta_x$, and the prefactor $\zeta_x$ in $C_{xx}^{(2)}$ and $C_{xx}^{(3)}$ with $(t+q_y/q)^2$ and $(t-q_y/q)^2$, respectively, which directly reflects the tilt-induced asymmetry in the velocity operators. We note that the broadening $\Gamma_{\mathrm{I}}$ and $\Gamma_{\mathrm{II}}$ depend, in general, on both the momentum $\mathbf q$ and the energy $\varepsilon$; this dependence has been suppressed throughout for notational compactness.

\subsection{Vertex corrections}
The Kubo formula in Eq.~\eqref{eq:Kubo} represents the bare bubble approximation, where vertex corrections from multiple impurity scattering are neglected. In systems with time-reversal symmetry $H(\mathbf{k})=H^*(-\mathbf k)$, these corrections vanish identically~\cite{burkov2011topological, Murakami2004}. For our tilted Hamiltonian, however, the tilted term $tq_y \mathbb{I}_4$ breaks this symmetry, necessitating a careful examination of these corrections.

\begin{figure}[t]
    \includegraphics[width=0.95\linewidth]{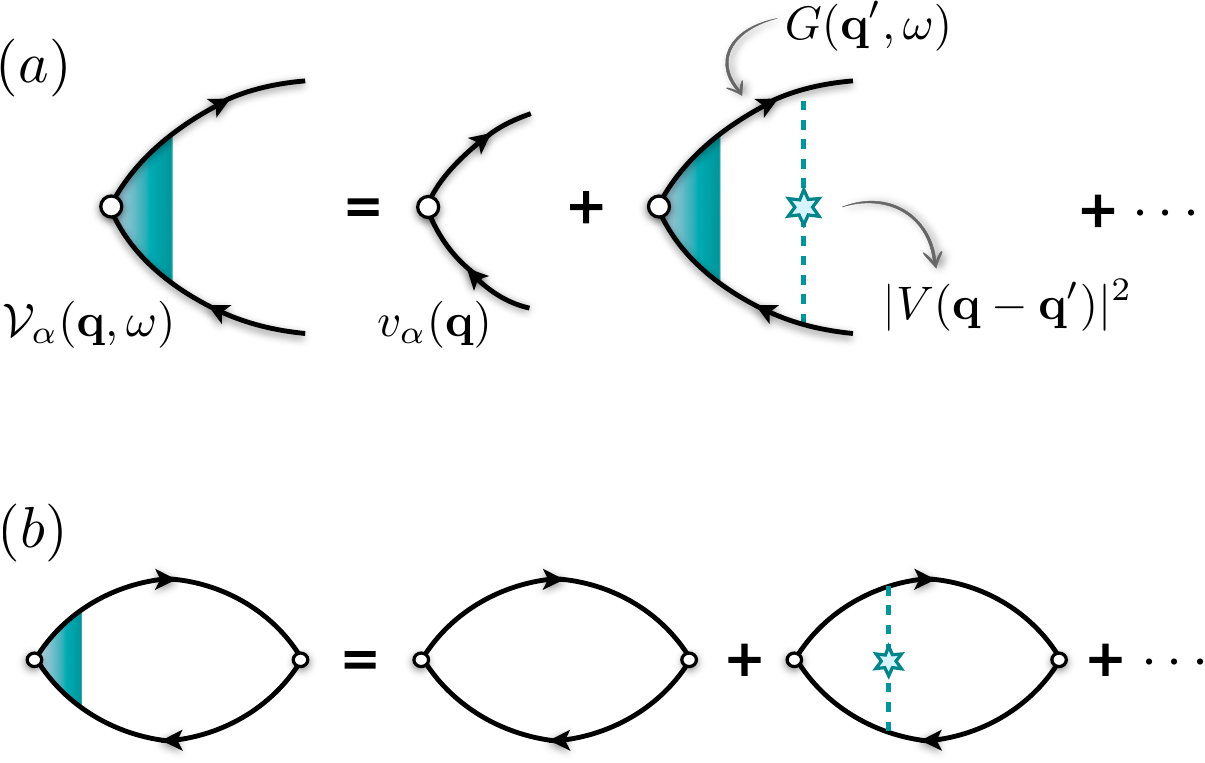}
	\caption{(a) Diagrammatic representation of the Bethe-Salpeter equation for the renormalized vertex function $\mathcal{V}_\alpha(\mathbf q,\omega)$ within the first Born approximation, as given by Eq.~\eqref{vcorrection}. (b) Corresponding substitution rules for the Kubo conductivity diagram.}
	\label{fig:1}
\end{figure}

We calculate analytically the first-order corrections within the first Born approximation at a fixed energy $\omega$. The renormalized vertex function $\mathcal{V}_\alpha(\mathbf q, \omega)$ satisfies the Bethe-Salpeter equation,
\begin{align}
    \mathcal{V}_\alpha(\mathbf q, \omega) & = v_\alpha (\mathbf{q}) + n_{\rm imp} 
     \int \frac{d^2 q'}{(2\pi)^2} |V(\mathbf{q}-\mathbf{q}')|^2 
    \nonumber \\
& \qquad  \times  G^R (\mathbf{q}',\omega) \mathcal{V}_\alpha (\mathbf{q}',\omega) G^A (\mathbf{q}',\omega) ,
    \label{vcorrection}
\end{align}
which is illustrated in Fig. \ref{fig:1} (a). Its corresponding substitutions for the Kubo formula diagram are also shown in Fig.  \ref{fig:1} (b). Depending on  the microscopic nature of the scattering potential (short-range vs. long-range Coulomb), the interaction kernel in Eq.~\eqref{vcorrection} behaves differently. However, as detailed in Appendix~\ref{app:vertex} for short-range defects and Appendix~\ref{app:vertex_long_range} for long-range Coulomb impurities, assuming small Fermi energies ($\varepsilon_F \ll \Lambda$), , the resulting correction to the longitudinal conductivity within the first Born approximation remarkably factors into a simple geometric scaling in both cases:
\begin{align}
    \label{c_correction}
    \sigma_{xx}(\varepsilon_F)\approx\sigma_{xx}^{(0)} (\varepsilon_F)\: \big[ 1+ \mathcal K({t}) \big],
\end{align}
where $\sigma_{xx}^{(0)}$ denotes the bare bubble conductivity, and $\mathcal K(t)$ is the vertex normalization factor modifying the bare velocity as 
$\mathcal{V}_x \approx v_x \big[1 + \mathcal{K}(t) \big]$. While the exact functional form of $\mathcal{K}(t)$ depends on the impurity type, it acts as a purely geometric renormalization dictated by the tilt parameter.

\section{Short-range impurity scattering potential}\label{sec:shortrange}
For isotropic scattering of electrons by short-ranged impurities, the potential takes the form:
\begin{align}
V(\mathbf{r})=V_0\sum_j\delta(\mathbf{r}-\mathbf{P}_j),
\end{align}
where $V_{0}$ is the scattering strength and $\{\mathbf{P}_j\}$ are randomly-distributed impurity locations. Throughout this work, we consider a dilute concentration of impurities such that multiple scattering events can be neglected, and the first-order Born approximation provides an adequate description.

\subsection{Subcritical conductivity($t<1$)} 
\label{sec:short_tless1}
Within the first-order Born approximation, the imaginary part of the self-energy (the inverse quasiparticle lifetime) is obtained by integrating over the Fermi surface. For $t<1$, the Fermi surface is closed and the density of states at the Fermi level remains finite. This leads to the following compact expressions for the damping rates:
\begin{equation}
\label{eq.gamma1}
	\Gamma_{\rm I,\rm II}=-\gamma_{\rm imp} \: \frac{|\omega|\,\Theta(\pm\omega)}{(1-t^2)^{\frac{3}{2}}}	 ,
\end{equation}
where $\Gamma_{\text{I}}$ corresponds to the conduction band ($\omega>0$) and $\Gamma_{\text{II}}$ to the valence band  ($\omega<0$),
and parameter $\gamma_{\rm imp}=n_{\rm imp} V^2_{0}/2\pi$ 
determines the strength of impurity broadening function.
The Heaviside step functions $\Theta(\omega)$ and $\Theta(-\omega)$ enforce that at zero temperature, only one band is occupied at a given energy $\omega$. Consequently, interband transitions between the valence and conduction bands are completely suppressed by the Pauli exclusion principle, and the corresponding interband term $C _{xx}^{(1)}$ in Eq.~\eqref{eq:Cxx} vanishes identically. The only remaining contributions are from intraband processes, described by $C _{xx}^{(2)}$ and $C_{xx}^{(3)}$, which involve scattering within the same band. The factor $(1-t^2)^{-3/2}$ in broadening functions reflects the tilt-induced modification of the density of states, which will be reinterpreted geometrically in Sec.~\ref{sec:geometric}.

To evaluate the conductivity integral in Eq.~\eqref{eq:Kubo}, it is convenient to transform to polar coordinates $(q,\theta)$ defined by $q_x=q\cos\theta$ and $q_y=q\sin\theta$. In the subcritical regime $t<1$ and for $\omega>0$, after performing the radial integration analytically, the longitudinal conductivity reduces to:
\begin{align}\label{t<1-old}
\sigma _{xx}&=\frac{-e^2}{4\pi^3}  \int  d\theta\,
{\cal J}_t(\theta)
\nonumber\\ 
& \times
\bigg[\frac{\Gamma_{\text{I}}^2+\varepsilon_F \Delta_q}{ \Gamma_{\text{I}}^2 + \Delta_q^2}+ \frac{\varepsilon_F}{ \Gamma_{\text{I}}}
\tan^{-1}\Big(\frac{\Delta_q}{\Gamma_{\text{I}}}\Big)
\bigg]_{q=0}^{q=q_f},
\end{align}
where $\Delta_q=\varepsilon_F\,-(1+t\sin\theta)q$ and
\begin{align}
  {\cal J}_t(\theta) =  \big( \frac{\cos \theta}{1+t\sin\theta} \big)^2.
\end{align}
Assuming large momentum cutoff $q_f \to \infty$, the above expression simplifies to
\begin{align}\label{t<1}
\sigma _{xx}=\frac{e^2}{4\pi^3}  \int  d\theta  {\cal J}_t(\theta)
\Big[
1+ \frac{\varepsilon_F}{\Gamma_I}
\Big(
\frac{\pi}{2}+
\tan^{-1}\big(\frac{\varepsilon_F}{\Gamma_{\text{I}}}\big)
\Big)\Big]
,
\end{align}

whose energy dependence completely disappears by noticing that $\Gamma_{\rm I}$ is linear in energy
and thus $\varepsilon_F/\Gamma_{\rm I}\equiv 
\gamma_{\rm imp}^{-1}(1-t^2)^{\frac{3}{2}} \Theta(\varepsilon_F)
$ does not depend on energy.
Therefore, in this case conductivity is solely governed by the geometric tilt parameter $t$, and the disorder strength factor $\gamma_{\rm imp}$. 

Physically, the bracketed term in Eq.~\eqref{t<1} reveals a profound separation of transport mechanisms. It consists of a purely quantum mechanical contribution (represented by the constant ``1") which survives in even strongly disordered limit, characterizing the intrinsic minimum conductivity of Dirac fermions. The second term, which is proportional to $\varepsilon_F/\Gamma_{\text{I}} \propto \tau_{\rm imp}$ (where $\tau_{\rm imp} = 1/\gamma_{\rm imp}$ is the scattering time), represents the semiclassical Drude-like transport governed by the finite quasiparticle lifetime.

\begin{figure}[t]
    \includegraphics[width=1 \linewidth]{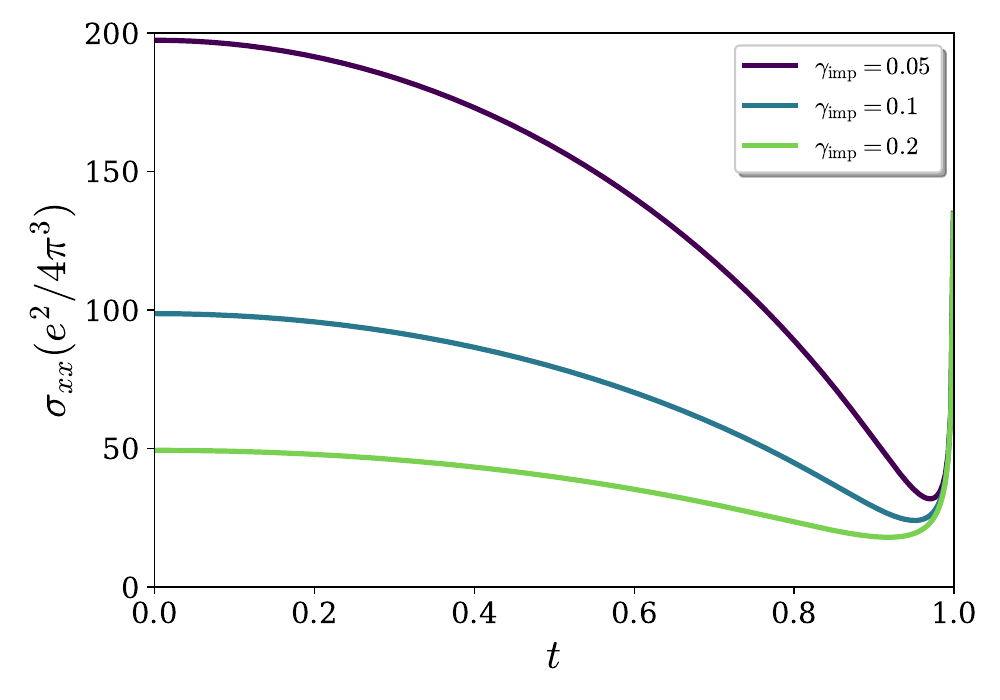}
	\caption{Subcritical longitudinal conductivity ($t<1$) for short-range impurities as a function of the tilt parameter $t$. The curves are plotted for different values of the dimensionless impurity scattering strength $\gamma_{\rm imp}=0.05$, $0.1$, and $0.2$, demonstrating a monotonic decrease of the macroscopic conductivity as the system approaches the Lifshitz transition.}
	\label{Sh_t_Less_1}
\end{figure}

Figure~\ref{Sh_t_Less_1} shows the conductivity calculated from Eq.~\eqref{t<1} for three values of the dimensionless scattering parameter, $\gamma_{\rm imp}=0.05$, $0.1$, and $0.2$, as a function of the tilt parameter $t$.
As the tilt increases, the ratio $\varepsilon_F/\Gamma_{\rm I}$ decreases monotonically, and thus the overall magnitude of the conductivity. However, in the vicinity of the critical tilt, $t \to 1$, the conductivity begins to increase sharply and exhibits an almost abrupt overshoot.


The overall decrease can be understood as a consequence of the geometric deformation of the Fermi contour, whereas the overshoot mainly originates from the radial integral $\int d\theta\, {\cal J}_t(\theta)$, which becomes very large as $t \to 1$.
It should be noted that the divergent behavior of $\sigma_{xx}$ 
as $t\to 1$ reflects the fact that at the critical point $t=1$, the system develops open Fermi lines. This necessitates the introduction of a finite but large momentum cutoff $\Lambda$ in order to obtain physically meaningful results. In the following subsections, where we consider both the critical and overcritical regimes characterized by open Fermi lines, we will therefore retain a cutoff $\Lambda$ to regularize the expressions and avoid nonphysical divergences.
In addition, we observe an almost linear dependence of the conductivity on the scattering time, $\tau_{\rm imp}=1/\gamma_{\rm imp}$, for all tilt values except near the critical tilt, $t \sim 1$. This is the expected behavior for short-range impurity scattering.

In the limit of an upright Dirac cone ($t=0$), our results reduce to the well-established behavior of disordered graphene with short-range impurities. Within the first-order Born approximation, the imaginary part of the self-energy scales linearly with energy, $\Gamma \propto \gamma_{\rm imp} |\omega|$, in agreement with earlier theoretical studies based on both semiclassical and diagrammatic approaches, including those by Ref.~\cite{DasSarma2007,DasSarma2008, Ando2002}. Consequently, the longitudinal conductivity $\sigma_{xx}$ exhibits a weak dependence on energy and approaches an approximately constant value, consistent with the standard Boltzmann transport results summarized by Ref.~\cite{DasSarma2011}. Furthermore, the suppression of interband contributions in the DC limit, reflected in the vanishing of $C_{xx}^{(1)}$, is also in line with previous Kubo-formula-based analyzes of short-range disorder~\cite{DasSarma2008}.

Building on these established results, our study generalizes the conductivity to the case of a tilted Dirac spectrum ($0 \le t<1$), where the tilt enhances the effective density of states by a factor $(1-t^2)^{-3/2}$, leading to an increased scattering rate and a corresponding monotonic reduction of $\sigma_{xx}$ as $t$ approaches unity.  This behavior is consistent with the general principle that stronger disorder or enhanced phase space for scattering suppresses conductivity \cite{Ostrovsky2006}, while highlighting the distinct role of band-structure tilt as an intrinsic mechanism for conductivity renormalization.

\subsection{Overcritical conductivity ($t>1$)} 
The energy-independent conductivity we have found for $t<1$ with short-range impurities resembles the point-defect-dominated regime in graphene ~\cite{DasSarma2007}. However, for $t>1$, the conductivity shows a nontrivial dependence on both Fermi energy $\varepsilon_F$ and momentum cutoff $\Lambda$, indicating that the overcritical regime is fundamentally distinct from any transport regime observed in untilted Dirac systems. In this regime, the strong tilt ($t>1$) leads to a Lifshitz transition, altering the Fermi surface topology from a closed elliptic contour to open hyperbolas. Crucially, this overcritical tilt pushes the valence band across the positive energy level $\omega>0$, creating coexisting electron and hole pockets (characteristic of Type II Dirac semimetals). Consequently both the conduction and valence bands scattering rates ($\Gamma_{\rm I,II}$) acquire a cutoff dependence.

By applying a momentum cutoff $\Lambda$ along the transverse direction ($|q_x| \le \Lambda$) to regularize the divergent result, the exact analytical integration of the broadening function yields a closed-form expression:
\begin{align}
\label{eq:self_energies_overcritical}
    \Gamma_{\text{I,II}} = -\gamma_{\rm imp} \bigg[ \frac{t \Lambda}{g} \pm \frac{\omega}{g^{3 \over 2}} 
    \sinh^{-1}
    \Big( \sqrt{g}\,\frac{\Lambda}{\omega} \Big) \bigg]\approx
    -\gamma_{\rm imp} \frac{t \Lambda}{g} 
    ,
\end{align}
where we have defined  $g=t^2-1$ for the overcritical regime,
and the approximate form holds when $\omega\ll \Lambda$. 
Here, the positive sign corresponds to the valence band ($\Gamma_{\text{I}}$) and the minus sign corresponds to the conductance band ($\Gamma_{\text{II}}$). 

Now in order to evaluate conductivity using expressions in Eqs. \eqref{eq:conductivity_integrand} and \eqref{eq:Cxx}, we note that the unperturbed energy dispersions $\xi_{\pm} = \omega - t q_y \pm q$ vanish along extended, ridge-like contours that asymptotically approach the directions $\sin\theta = \pm 1/t$ for $q\to\infty$. Hence, these open contours extend to infinity, this would lead to divergent conductivity integrals.
In fact, the divergences of the 
level broadening and also conductivity integrals with the cutoff $\Lambda \to \infty$ is a characteristic manifestation of an ultraviolet divergence inherent in continuum field theories. In this framework, the dispersion relation is assumed to remain valid at arbitrarily high momenta, allowing an infinite density of states to contribute to the result. In realistic condensed matter systems, this divergence is physically regularized by the underlying crystal lattice. The existence of a finite Brillouin Zone imposes a natural momentum cutoff $\Lambda \approx \pi/a$ (where $a$ is the lattice constant), and the periodic nature of the lattice potential ensures a bounded bandwidth. Consequently, the lattice provides a self-consistent regularization that avoids the \emph{ultraviolet catastrophe} of the continuum approximation, ensuring that physical observables remain finite.

\begin{figure}[t]
 \includegraphics[width=0.95\linewidth]{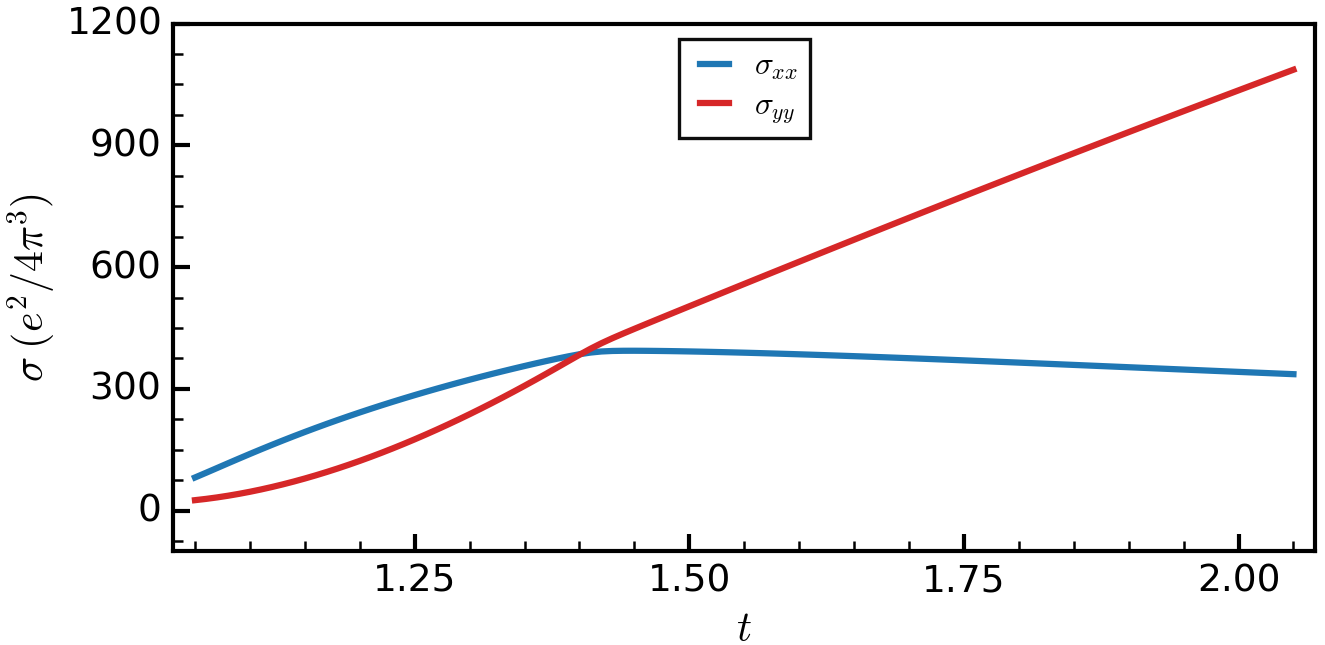}
    \caption{Transport anisotropy in the overcritical regime ($t > 1$) for short-range impurities. The conductivity perpendicular to the tilt direction, $\sigma_{xx}$ (blue curve), exhibits a non-monotonic behavior, peaking near $t \approx \sqrt{2}$ and decreasing at larger tilts. In contrast, the conductivity along the tilt axis, $\sigma_{yy}$ (red curve), increases monotonically without bound, reflecting profound transport anisotropy. Results are shown for a scattering strength $\gamma_{\rm imp} = 0.1$ and a small Fermi energy relative to the momentum cutoff ($\varepsilon_F/\Lambda = 2.5 \times 10^{-3}$).}
    \label{fig:anisotropy}
\end{figure}

To maintain consistency of the theoretical formulation, we evaluate the conductivity integral in Eq.~\eqref{eq:conductivity_integrand} using the same momentum cutoff $\Lambda$ employed for the self-energies. The numerical results for $\sigma_{xx}$ and $\sigma_{yy}$ in the overcritical regime are presented in Fig.~\ref{fig:anisotropy}. Particularly, $\sigma_{xx}$, the conductivity perpendicular to the tilt axis exhibits non-monotonic behavior, reaching a peak near $t \approx \sqrt{2}$ before decaying as $1/t$.


This unusual behavior can be understood by analyzing the interplay between the anisotropic velocities and the available phase space on the hyperbolic Fermi surface. As discussed in Appendix \ref{app:anisotropy}, the group velocities in the tilt direction grows as $v_y \sim t-1/t$, while the transverse velocity saturates as $v_x \sim \sqrt{t^2-1}/t \to 1$ by increasing the tilt.  
Assuming small energies and working in an on-shell approximation $\varepsilon_{\pm}({\bf q})\sim 0$, the asymptotic behavior of hyperbolic Fermi surface reads $q_{x}=\pm \sqrt{t^2-1}q_y$. Therefore, due to this elongation, the effective bound for the transverse momentum $q_x$ scales as $\Lambda_x \propto \Lambda \sqrt{t^2-1}/t$.
The interplay of these effects 
of tilt on effective velocities and the bounds of transverse momentum, leads to a qualitative scaling form for the conductivity $\sigma_{xx}$ as
\begin{align}
    \sigma_{xx}\sim \int d^2q \:\delta\big[\varepsilon({\bf q})\big] \, v_x^2({\bf q}) \sim 2\Lambda \frac{t^2-1}{t^2},
\end{align}
as detailed in Appendix \ref{app:anisotropy}.

We further explore the conductivity component $\sigma_{yy}$ and compare it with 
$\sigma_{xx}$, the results of which are shown in Fig. \ref{fig:anisotropy}.
In contrast to the transverse transport (along $x$), the conductivity parallel to the tilt direction ($y$) grows unboundedly with $t$, revealing a large transport anisotropy of Type II Weyl/Dirac semimetals.
This profound anisotropy naturally emerges from the interplay between the group velocities and the restricted phase space on the hyperbolic Fermi surface as function of tilt parameter $t$.
Employing the same analysis (see Appendix \ref{app:anisotropy}), 
we find that the leading behavior
for conductivity along tilt direction follows $\sigma_{yy} \propto (t^2-1)^{3/2}/t^2$, 
in qualitative agreement with the numerical results in Fig. \ref{fig:anisotropy}.

\subsection{Critical conductivity ($t=1$)}

\begin{figure}[t]
	\includegraphics[width=0.98\linewidth]{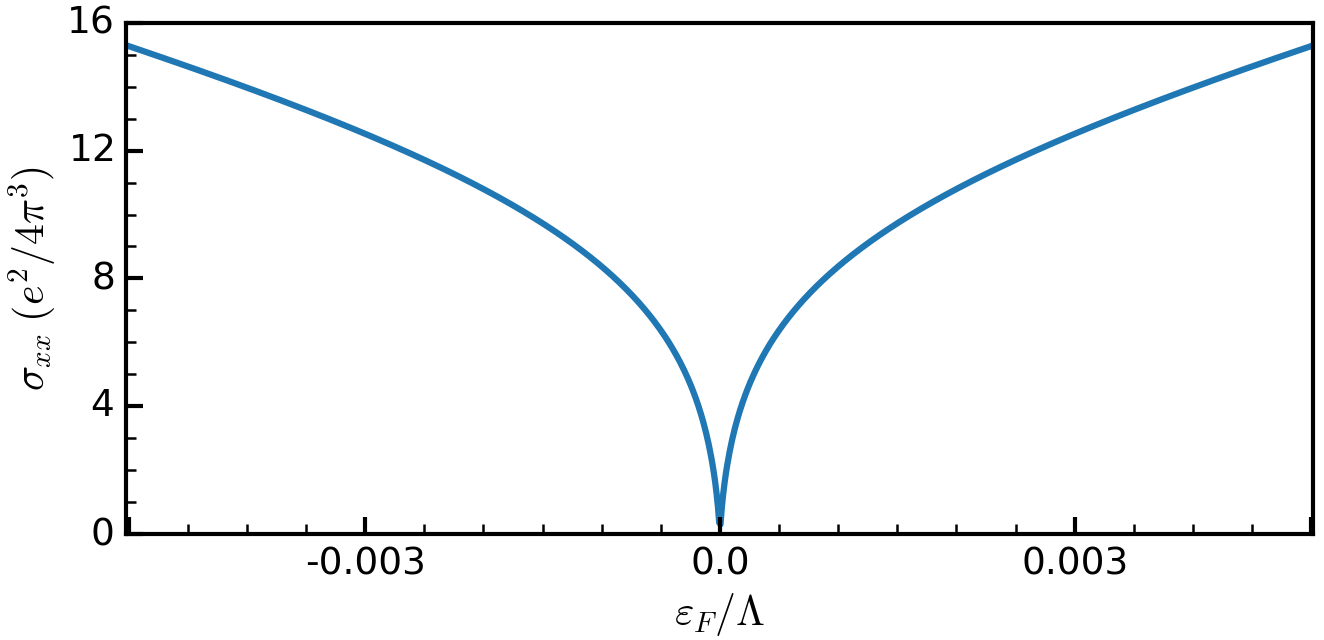}
	\caption{DC conductivity at the exact critical point ($t=1$) for short-range impurities as a function of energy $\varepsilon_F$ near the Dirac point ($\varepsilon_F/\Lambda \in[-0.005, 0.005]$). The vanishing conductivity as $\varepsilon_F \to 0$ is a direct macroscopic consequence of the $1/\sqrt{|\omega|}$ divergence in the scattering self-energy.}
	\label{fig:Short_Critical}
\end{figure}

At the exact Lifshitz transition ($t=1$), the nature of the scattering changes drastically. For a finite cutoff $\Lambda$ and $\omega \ll \Lambda$, the exact analytical evaluation of the self-energies results in
\begin{align}
    \Gamma_{\rm I, II}= \gamma_{\rm imp}  \Theta(\pm\omega) \left[ \frac{(\Lambda + 2|\omega|)\sqrt{|\omega|(|\omega| + 2\Lambda)}}{-3|\omega|} \right].
\end{align}
In the low-energy limit ($\omega \ll \Lambda$), the self-energies display a strong divergent behavior proportional to $\Lambda^{3/2}/\sqrt{|\omega|}$. This singularity deeply affects the conductivity. As depicted in Fig.~\ref{fig:Short_Critical}, numerical integration of the Kubo formula at $t=1$ shows a highly localized dip in conductivity around the Dirac point. The divergence in the scattering rate effectively strongly suppresses the transport, marking a clear transport signature of the critical topological transition.

\subsection{Vertex corrections to the conductivity}

As shown in Appendix \ref{app:vertex}, the vertex corrections for $t<1$ are given by an energy-independent, purely geometric renormalization factor $\mathcal{K}(t) = (1 - t^2)/(1 + \sqrt{1 - t^2})$, which rescales the conductivity without altering its frequency dependence. For $t>1$, the situation initially appears more intricate due to the open hyperbolic topology of the Fermi surface. In this regime, both the bare vertex integral and the scattering self-energy diverge linearly with the ultraviolet momentum cutoff $\Lambda$. However, these divergences cancel exactly in the physical limit ($\Lambda \gg \omega$). Consequently, the first-order vertex renormalization factor again reduces to a purely geometric form,
$\mathcal{K}(t) = 2(t^2 - 1)/t^2$
,
which is governed solely by the tilt parameter $t$ and contains no explicit dependence on $\Lambda$ or $\omega$.

At the critical point $t = 1$, our analytical calculation shows that the first-order vertex correction vanishes identically for both short-range and long-range impurity potentials. Consequently, $\mathcal{K}(t=1) = 0$ (so that the renormalized vertex satisfies $\mathcal{V}_x = v_x$), and the bare-bubble approximation becomes exact to first order in the impurity concentration. This result is particularly noteworthy, as it indicates that vertex corrections do not contribute at the Lifshitz transition due to the diverging density of states, thereby greatly simplifying the theoretical analysis of the critical regime.

\begin{figure}[t]
	\includegraphics[width=0.95\linewidth]{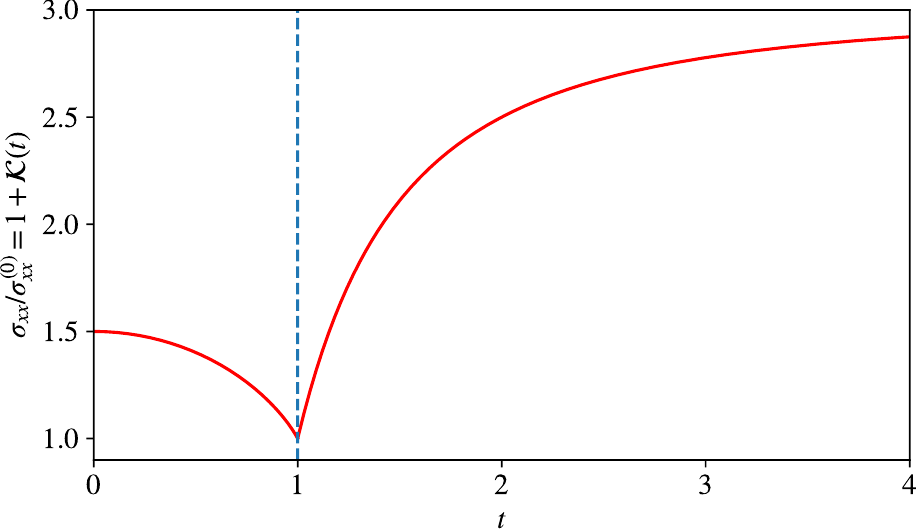}
	\caption{The effect of the first-order vertex correction on the longitudinal conductivity for short-range impurities, represented by the ratio  $\sigma_{xx} / \sigma_{xx}^{(0)} = 1 + \mathcal{K}(t)$. The geometric correction factor strongly renormalizes the conductivity in the subcritical regime $t<1$, vanishes identically at the critical point ($t=1$), and scales proportionally with $t$
 in the overcritical regime($t>1$).}
	\label{fig:Vertex_Correction}
\end{figure}

The role of vertex corrections and the resulting velocity renormalization is illustrated in Fig.~\ref{fig:Vertex_Correction}, which shows the full conductivity including the vertex factor $\mathcal{K}(t)$ compared to the bare-bubble conductivity $\sigma_{xx}^{(0)}$. In the subcritical regime, the vertex correction provides a substantial enhancement. As the tilt increases and the system approaches the Lifshitz transition ($t \to 1$), the ratio drops to unity, corresponding to the complete suppression of vertex corrections due to the diverging density of states. In the overcritical regime, the correction factor increases again with further overtilting ($t > 1$) and asymptotically approaches $3$ for very large tilt $t \gg 1$.

\subsection{Global conductivity behavior across the Lifshitz transition}

\begin{figure}[t]
	\centering
	\includegraphics[width=0.95\linewidth]{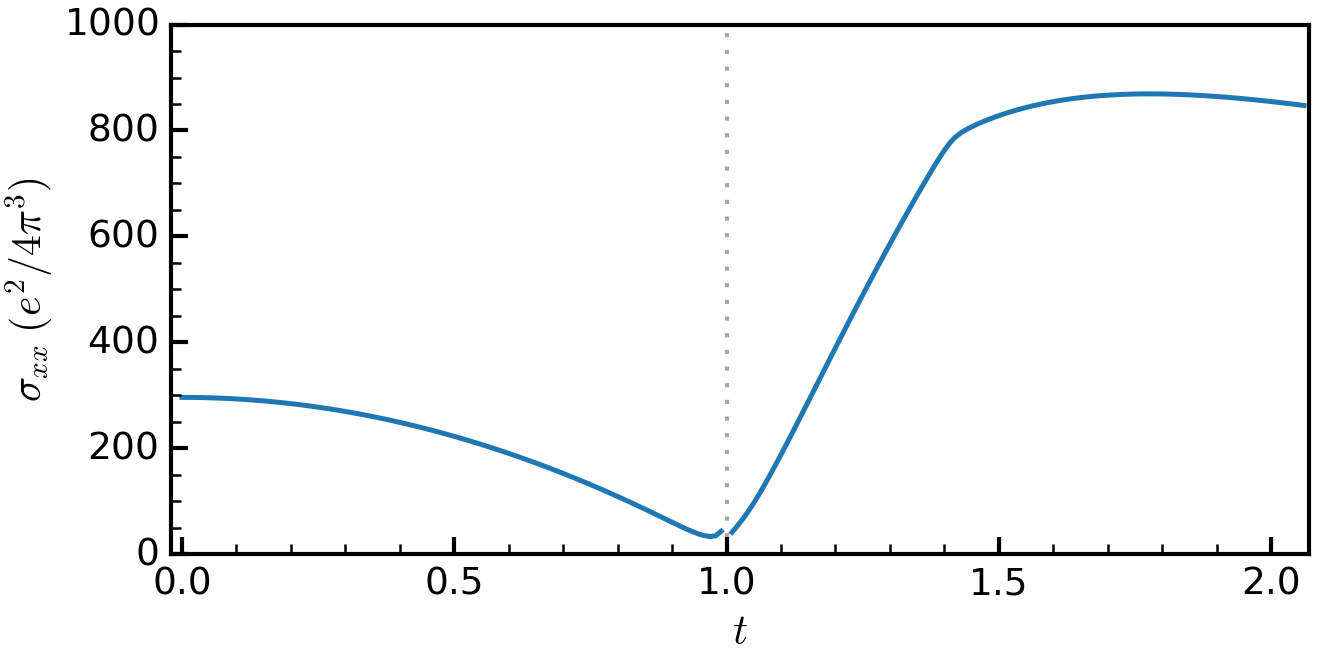}
	\caption{Global evolution of the longitudinal DC conductivity $\sigma_{xx}$ across the full tilt parameter space ($0 \le t \le 2$) for short-range impurities. The overarching curve smoothly connects the frequency-independent subcritical plateau ($t<1$), the singular conductivity dip at the Lifshitz transition ($t=1$), and the non-monotonic, cutoff-dependent transport peak in the overcritical Type II regime ($t>1$).}
 	\label{fig:Global_Conductivity}
 \end{figure}

Synthesizing the results from the subcritical, critical, and overcritical regimes provides a comprehensive global picture of how the tilt parameter dictates DC transport. Figure~\ref{fig:Global_Conductivity} displays the continuous evolution of the longitudinal conductivity $\sigma_{xx}$ across the full tilt parameter space $0 \le t \le 2$ at a fixed energy and cutoff.

 The global profile is characterized by three distinct phases: (i) A largely flat plateau in the untilted and weakly tilted regimes ($t \lesssim 0.4$), corresponding to the standard point-defect dominated transport where geometric modifications smoothly scale the bare conductivity. (ii) A sharp, localized suppression (dip) exactly at the Lifshitz transition ($t=1$), driven by the van Hove singularity in the density of states which critically enhances the scattering rate. It should be noted that the sharp visual discontinuity connecting the curve exactly at $t=1$ is a regularization artifact; it arises from joining the cutoff-independent analytical of the closed subcritical regime with the finite-cutoff numerical evaluation strictly required for regularizing the open Fermi topologies at $t\ge 1$.  (iii) A resurgence in the overcritical regime ($t>1$) where the transition to Type II Dirac fermions manifests as a non-monotonic bump near $t \approx \sqrt{2}$, eventually giving way to asymptotic decay due to phase-space constriction. This unified curve establishes the tilt parameter $t$ as a highly versatile tuning knob for engineering profound topological and transport phase transitions in Dirac materials.
 
\section{Long-range impurity scattering potential}\label{sec:longrange}

For charged impurities, the scattering potential is given by the screened Coulomb interaction. In real space, the potential of a single impurity located at $\mathbf{P}_j$ reads:
\begin{align}
V(\mathbf{r}) = \frac{e^2}{4\pi\epsilon} \sum_j \frac{1}{|\mathbf{r} - \mathbf{P}_j|},
\end{align}
where $\epsilon$ is the absolute permittivity of the surrounding medium. In two dimensions, the Fourier transform of the Coulomb potential takes the form
\begin{align}
V(\mathbf{q}) = \frac{e^2}{2\epsilon} \frac{1}{q},
\end{align}
where $q = |\mathbf{q}|$. Throughout this section, we absorb the constants into a single dimensionless coupling strength $\alpha = e^2/(4\pi\epsilon v_F \hbar)$ (the fine-structure constant of the system). Now, the level broadening in first Born approximation due to the impurities for the band $\lambda$ reads
\begin{align}
    \Gamma_\lambda(\mathbf q , \omega)=-\pi n_{\rm imp} \int \frac{d^2 \mathbf{q}'}{(2\pi)^2} |V_{\mathbf q - \mathbf{q}'}|^2
    \delta(\omega-\varepsilon^{(\lambda)}_{\mathbf{q}'}).
\end{align}

\subsection{Subcritical conductivity($t<1$)}
\label{sec:long_tlt1}

\begin{figure}[t]
  \includegraphics[width=0.99\linewidth]{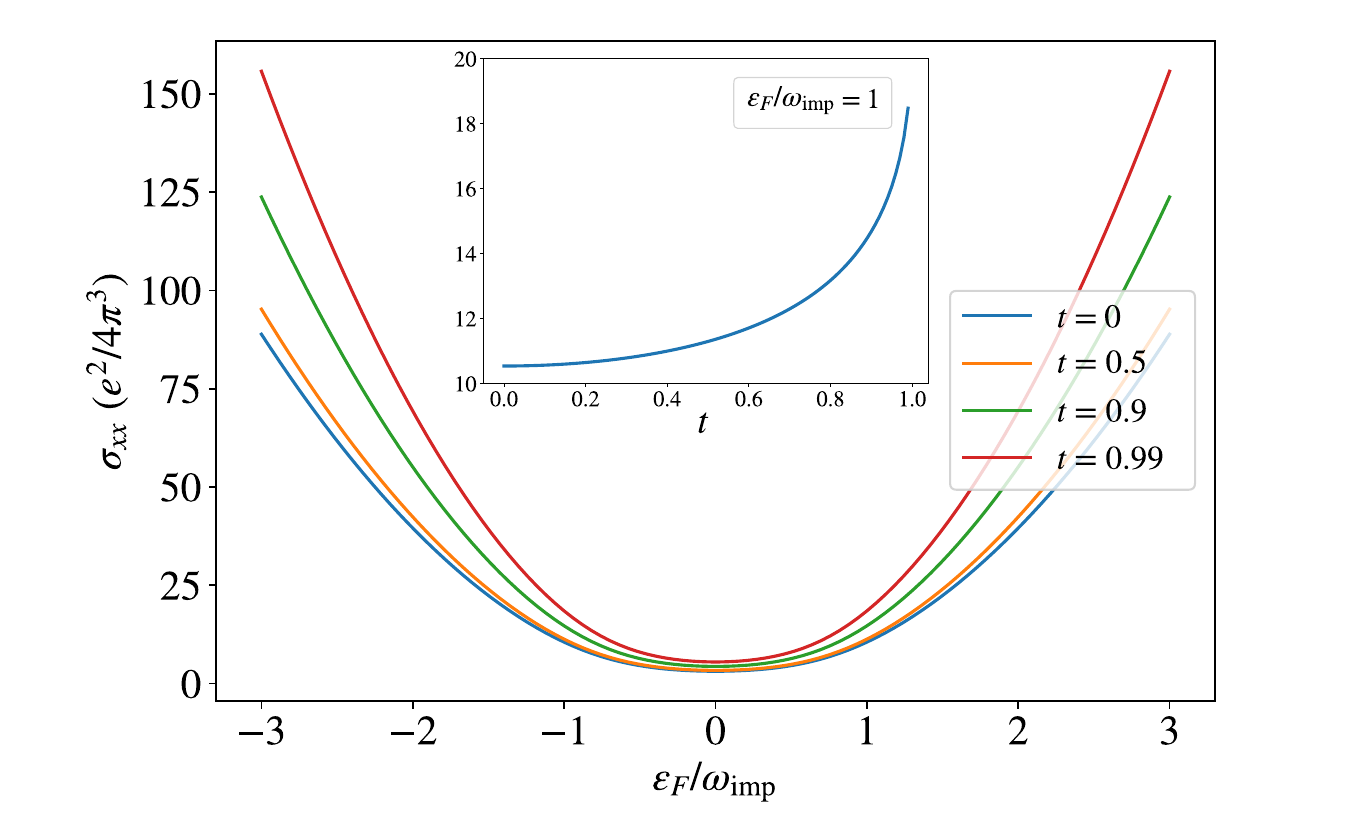}
    \caption{Subcritical longitudinal conductivity ($t<1$) for long-range Coulomb scattering as a function of the scaled Fermi energy $\varepsilon_F/\omega_{\rm imp}$. In sharp contrast to the short-range case, the conductivity exhibits a strong, nearly quadratic energy dependence. The inset illustrates that, at a fixed finite energy, the conductivity is enhanced as the tilt parameter approaches the critical value ($t\rightarrow 1$).}\label{fig:9}
\end{figure}

Unlike the short-range impurity case, the momentum dependence of the Coulomb potential $V(q)\propto 1/q$ introduces a non-trivial angular integral that cannot be trivially factorized. However, a detailed calculation yields the following expressions for the self-energies in the subcritical regime at zero external momentum $\mathbf{q}=0$: 
\begin{align}\label{eq:longrange_sub_Gamma}
	\Gamma_{\text{I,II}}\sim-\frac{\pi^2 n_{\rm imp}\alpha^2}{2}\frac{\Theta(\pm \omega)}{|\omega|} 
    =
    -\frac{\omega_{\rm imp}^2}{|\omega|} \, \Theta(\pm \omega),
\end{align} 
for the conduction and valence bands, respectively. Here, the disorder density and the dimensionless interaction constant
have been combined into a single energy scale $\omega_{\rm imp} = \pi \alpha \sqrt{n_{\rm imp}/2}$. In contrast to the short-range potential, evaluating the broadening function in this regime reveals an exact mathematical cancellation. The enhancement of the density of states along the tilt direction is perfectly balanced by the $1/q^2$ suppression of the Coulomb scattering matrix elements. Consequently, the self-energies in the subcritical regime become independent of the tilt parameter $t$.

Using the tilt-independent self-energies derived for the long-range Coulomb potential in Eq. \eqref{eq:longrange_sub_Gamma}, the conductivity follows the same formal expression as Eq. \eqref{t<1}. However, there is a crucial difference: the energy dependence of the broadening function, $\Gamma_{\rm I} \propto 1/|\omega|$, leads to an energy dependence in the conductivity. As illustrated in Fig.~\ref{fig:9}, the conductivity exhibits a nearly quadratic variation with the Fermi energy, $\sigma_{xx} \propto (\varepsilon_F/\omega_{\rm imp})^2$. Interestingly, since the broadening function is independent of the tilt, the conductivity $\sigma_{xx}$ in the subcritical regime factors into a separable form: $\sigma_{xx} = f(\varepsilon_F)g(t)$. The variation with respect to the tilt, denoted by $g(t)$, is shown in the inset of Fig.~\ref{fig:9} for $\varepsilon_F/\omega_{\rm imp}=1$. Apart from a numerical prefactor, this dependence remains consistent across all energy scales.

\subsection{Overcritical conductivity ($t>1$)}
\label{sec:long_tgt1}

\begin{figure}[t]
	\includegraphics[width=0.99\linewidth]{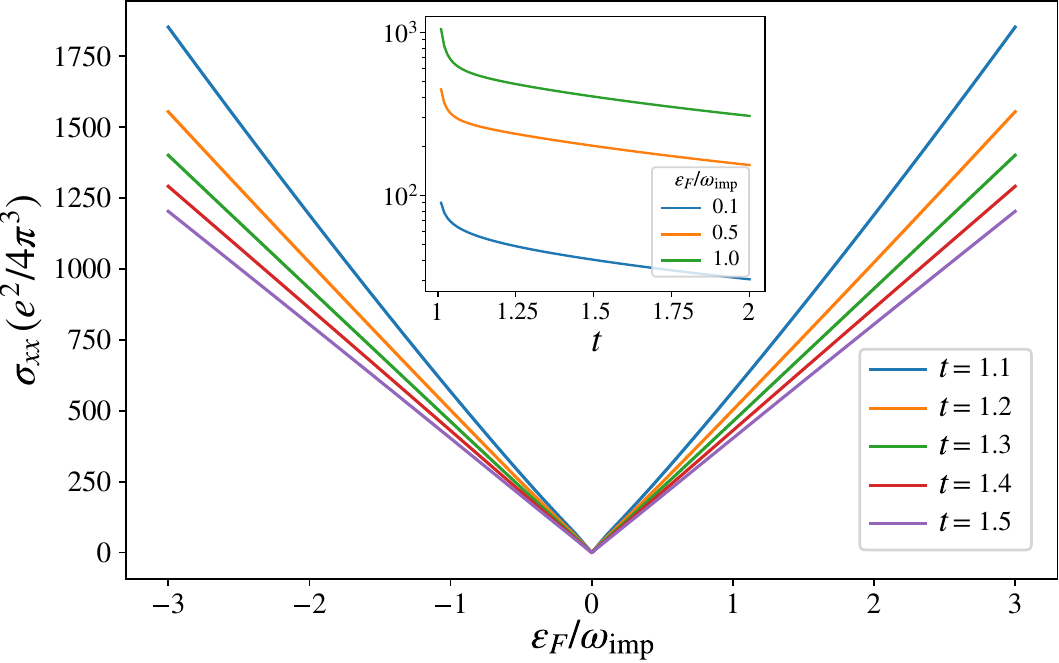}
	\caption{Overcritical conductivity for long-range Coulomb scattering as a function of the scaled Fermi energy 
$\varepsilon_F / \omega_{\text{imp}}$, calculated for a large ultraviolet cutoff 
$\Lambda / \omega_{\text{imp}} = 100$ and various tilt parameters $t$. The inset explicitly shows the continuous evolution of the conductivity with tilt at a fixed energy, highlighting a pronounced macroscopic enhancement near the Lifshitz transition ($t \to 1^{+}$) followed by a monotonic decrease for larger tilts.}
	\label{fig:Long_Overcritical_w}
\end{figure}

In the overcritical regime, the Lifshitz transition fundamentally alters the Fermi surface
topology, transforming closed ellipses into open hyperbolas. As a result, both the conduction
and valence bands extend across the full energy spectrum ($\omega \in \mathbb{R}$), allowing
interband transitions to contribute at all energies.

In sharp contrast to the short-range case, the long-range Coulomb potential suppresses
scattering at large momenta as $1/q^2$, which naturally eliminates ultraviolet divergences in both the self-energies. Consequently, the level broadening in the overcritical regime with long-range scattering is intrinsically  UV-finite as the suppression of the Coulomb potential in large momentum naturally regularize the scattering rate expression. However, evaluating the macroscopic conductivity via the Kubo formula still necessitates an explicit momentum cutoff $\Lambda$; because the open hyperbolic Fermi lines extend to infinity, the transport integrals would otherwise develop a logarithmic divergence.

The level broadening functions corresponding to the two band dispersions
$\epsilon_{\pm}(\mathbf{k})$ take the form
\begin{align} \label{eq:Gamma-longrange-t>1}
    \Gamma_{\rm I, II} = -\frac{2\pi n_{\rm imp}\alpha^2}{|\omega|}
    \sec^{-1}\!\left(\mp\,{\rm sgn}(\omega)\, t\right).
\end{align}
Compared to the subcritical regime, the key difference is a mild dependence on the tilt
parameter entering through $\sec^{-1}(\pm t) \equiv \arccos(\pm 1/t)$, which varies smoothly
between $0$ and~$\pi$.

Using these scattering rates, the conductivity is evaluated numerically via the Kubo formula
given in Eq.~\eqref{eq:conductivity_integrand}. The results are shown in
Fig.~\ref{fig:Long_Overcritical_w} for several values of the tilt parameter, as a function
of the scaled Fermi energy $\varepsilon_F/\omega_{\rm imp}$, at a fixed ultraviolet cutoff
$\Lambda = 100\,\Omega_{\rm imp}$. We note that, as in the short-range overcritical case,
the open Fermi lines extend to infinity and a finite cutoff is still required to regularize
ultraviolet divergences, even though long-range scattering renders the self-energy itself
UV-finite.

In the overcritical regime, the density of states is nearly constant in energy. Combined with
a scattering time that scales linearly with energy, this yields a conductivity that grows
approximately linearly with Fermi energy, as seen in Fig.~\ref{fig:Long_Overcritical_w}.
This contrasts with the nearly quadratic behavior observed in the subcritical case, where the
density of states itself scales linearly with energy; the combined enhancement from both the
density of states and the scattering time produces the quadratic energy dependence there.

Furthermore, as shown in the inset of Fig.~\ref{fig:Long_Overcritical_w}, the conductivity
increases as the tilt parameter decreases toward the critical value $t \to 1$. This
enhancement is primarily geometric in origin, arising from the divergence of the group
velocity and the associated growth of the density of states as the system approaches the
Lifshitz transition. This effect is discussed in further detail in
Sec.~\ref{subsec:global_longrange}.

\subsection{Critical conductivity ($t=1$)}

\begin{figure}[t]
	\includegraphics[width=0.99\linewidth]{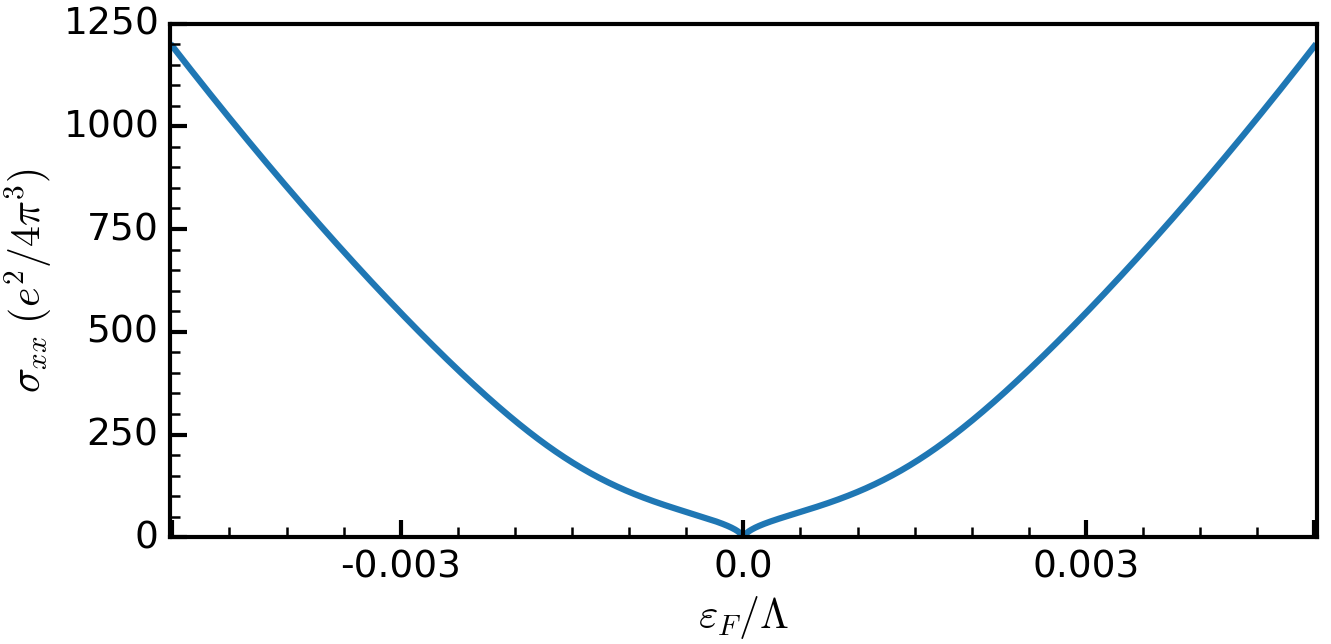}
	\caption{Long-range conductivity exactly at the critical point ($t=1$) as a function of energy near the Dirac point ($\varepsilon_F/\Lambda \in [-0.004, 0.004]$). Similar to the short-range scenario, the intense singularity in the critical self-energy causes a sharp, highly localized dip in conductivity precisely at the Dirac point ($\varepsilon_F = 0$).}
	\label{fig:Long_Critical}
\end{figure}

At the Lifshitz transition point for long-range Coulomb impurities, the self-energies adopt a distinct inverse-energy scaling modulated by arc-tangent functions:
{\begin{equation}
	\Gamma_{\rm I,\rm II} =  -\frac{n_{\rm imp}\alpha^2}{\pi} \Theta(\mp\omega) \left[ \frac{\tan^{-1}\sqrt{\frac{\omega \mp 2\Lambda}{\omega}}}{|\omega|} \right].
\end{equation}
Similar to the short-range critical case, this strong energy dependence in the denominator drives the conductivity to display sharp features near the Dirac point. As plotted in Fig.~\ref{fig:Long_Critical}, a highly localized plunge in the conductivity is observed in the extremely narrow window around $\omega=0$, reaffirming that the $t=1$ critical state robustly suppresses DC transport regardless of the scattering potential range.

It is crucial to distinguish this energy-dependent suppression precisely at the Dirac point ($\omega=0$) from the macroscopic tilt-dependent behavior. As will be discussed in Sec.~\ref{subsec:global_longrange} , for a fixed finite Fermi energy, the geometric enhancement of the density of states as $t\rightarrow 1$ ultimately drives a pronounced peak in the conductivity. 

\subsection{Global conductivity behavior across the Lifshitz transition}\label{subsec:global_longrange}

As shown in Appendix~\ref{app:vertex_long_range}, in the presence of long-range impurities
and within the small external momentum approximation, the vertex correction in the first Born
regime remains essentially constant, ${\cal K} = {\rm cte}$, except in the immediate
vicinity of the Lifshitz transition $t \to 1$. As argued in detail in
Appendix~\ref{app:vertex_long_range}, the sharp divergence of the density of states drives
the Coulomb scattering self-energy to infinity. Consequently, the integral term in the
Bethe--Salpeter equation becomes heavily suppressed, causing the vertex correction to vanish
identically ($\mathcal{V}_\alpha = v_\alpha$). In other words, the bare bubble approximation
becomes exact to first order in the impurity concentration.

Since the vertex correction is nearly constant throughout the tilt parameter space and
vanishes precisely at the critical point, its overall effect is trivial. This means that the
global variation of the conductivity in the presence of long-range Coulomb impurities is
accurately captured by the bare bubble approximation across the entire range of tilt parameter
values.

The insets of Figs.~\ref{fig:9} and \ref{fig:Long_Overcritical_w} clearly show that the
conductivity becomes increasingly enhanced as the system approaches the Lifshitz transition
point $t \to 1$. While the short-range scattering rate diverges strongly at the Lifshitz
transition, the scattering rate in the long-range case exhibits little to no dependence on
the tilt. As a result, whereas in the short-range case the large scattering rate suppresses
the conductivity near the transition, in the long-range case the enhanced density of states,
combined with a moderate scattering rate, produces a pronounced enhancement of the
conductivity at the Lifshitz transition.
}

\section{Geometric Reformulation of the Tilted Dirac Cone} \label{sec:geometric}
The conventional treatment of Dirac cone tilt incorporates the tilt parameter $t$ directly as a modification to the Hamiltonian. However, a profound and physically illuminating alternative is to reinterpret the tilt not as a mere algebraic term in the Hamiltonian, but as the manifestation of an effective curvature in momentum space.

In the standard untilted case ($t=0$), the low-energy physics is governed by an isotropic Dirac cone, which corresponds to a flat, Euclidean momentum space with the metric $ds^2=dq_x^2+dq_y^2$. Introducing a tilt in the $q_y$-direction fundamentally alters this geometric structure. The system can be described by a new, effective metric:
\begin{align}
    ds^2 = dq_x^2 + (1-t^2) dq_y^2 = dq_x^2 + g \, dq_y^2
\end{align}
where $g = \det(\hat{g}) = 1-t^2$ is the determinant of the effective metric tensor $\hat{g} = \text{diag}(1, 1-t^2)$. This metric succinctly encodes the anisotropic distortion of the Fermi surface caused by the tilt. 

Most importantly, the velocity operators, which are fundamental to the Kubo formula, are no longer given by simple partial derivatives. Instead, they must be replaced by their covariant derivatives, which intrinsically account for the curvature of the momentum space:
\begin{align}
    v_x &= \frac{\partial H}{\partial q_x} \rightarrow \frac{\partial H}{\partial q_x} \nonumber \\
    v_y &= \frac{\partial H}{\partial q_y} \rightarrow \left( \frac{1}{g} \right) \frac{\partial H}{\partial q_y}
\end{align}
This geometric reformulation reveals the anisotropic nature of the tilt-induced effects on transport coefficients. Specifically, while the $x$-component of the velocity operator and consequently the diagonal conductivity $\sigma_{xx}$ remain unaffected by this particular metric transformation, the $y$-component transforms as $v_y \rightarrow v_y/g$. This directly impacts the computed values of $\sigma_{yy}$ and the off-diagonal Hall conductivity $\sigma_{xy}$, providing a natural geometric explanation for the anisotropic transport behavior observed in tilted Dirac systems.

Another significant consequence of this geometric picture arises in the evaluation of momentum-space integrals. Integrals of the form $\int d^2 q/(2\pi)^2$ must be modified to include the invariant volume element: $\int \sqrt{|g|} \, d^2 q/(2\pi)^2$. Furthermore, the energy constraint delta function in curved space transforms as $\delta(\omega-\sqrt{q_x^2+q_y^2}) \rightarrow \delta \bigg(\omega-\sqrt{q_x^2+q_y^2} \bigg)/\sqrt{|g|}$.

This geometric volume factor dictates the density of states and deeply influences the scattering lifetimes. For the subcritical regime ($t<1$), the imaginary part of the self-energy (scattering rate) is strongly renormalized by the curved space. Using the untilted broadening function $\Gamma_0 = \frac{n_{\rm imp} V^2_{0}}{2} \omega$, the effective self-energy becomes:
\begin{align} \label{Gamma_Geometric}
    \Gamma_{\text{II}} = \frac{\Gamma_0}{|g|^{3/2}} = \frac{n_{\rm imp} V^2_{0}}{2}\frac{\omega}{(1-t^2)^{\frac{3}{2}}}
\end{align}
This reveals that the $(1-t^2)^{3/2}$ term is purely a geometric manifestation of the modified phase-space volume.

\subsection{Analytical Conductivity in the Geometric Framework}
We now apply this geometric insight to the explicit calculation of the longitudinal conductivity at a constant energy $E=\omega > 0$. At zero temperature, the interband transitions are suppressed due to the Heaviside step functions, leading to $C _{xx}^{(1)} = C _{xx}^{(3)} = 0$. The longitudinal conductivity is governed entirely by intraband scattering ($C _{xx}^{(2)}$):
\begin{align}
    \sigma_{xx} = \frac{e^2}{\pi} \int \frac{d^2 q}{(2\pi)^2} \zeta_x \frac{2\Gamma_{\text{II}}^2}{\left(\Gamma_{\text{II}}^2+\xi_{-}^2\right)^2}
\end{align}
While this integral can be solved algebraically using standard polar coordinates, interpreting the final result through our effective metric $g$ yields a remarkably elegant relation. The geometric deformation maps the circular Fermi contour to an ellipse, reshaping both the density of states (captured by Eq.~\ref{Gamma_Geometric}) and the angular distribution of velocities. 

By performing the integration over the radial and angular momentum components analytically, the tilt parameter $t$ can be completely substituted by the metric determinant $g = 1-t^2$. The analytical result for the conductivity at a constant energy becomes:
\begin{align}
    \sigma_{xx} = \frac{e^2}{\pi n_{\rm imp} V^2_{0}} \left( \frac{g}{1-g} \right) \left( 1 - \sqrt{g} \right)
\end{align}
This analytical expression is profoundly insightful. It demonstrates that the longitudinal conductivity $\sigma_{xx}$—despite its velocity operator $v_x$ remaining geometrically unaffected by the covariant transformation—is still heavily renormalized. This renormalization stems entirely from the metric determinant $g$, which governs the geometry of the scattering phase space. 

As a sanity check, in the flat space limit where the tilt vanishes ($t \to 0 \implies g \to 1$), we can expand $\sqrt{g} \approx 1 - (1-g)/2$. The geometric factors gracefully cancel out:
\begin{align}
    \lim_{g \to 1} \sigma_{xx} = \frac{e^2}{\pi n_{\rm imp} V^2_{0}} \left( \frac{g}{1-g} \right) \left( \frac{1-g}{2} \right) = \frac{e^2}{2\pi n_{\rm imp} V^2_{0}}
\end{align}
which perfectly recovers the standard Drude conductivity for isotropic, untitled Dirac fermions, thereby verifying the geometric framework.

\section{Discussion}\label{sec:discussion}

We present a comprehensive theoretical investigation of the DC electronic conductivity in two-dimensional tilted Dirac semimetals, systematically mapping transport properties across the subcritical (Type I), critical, and overcritical (Type II) regimes. Using the Kubo formalism within the first-order Born approximation, we analyze the intricate interplay between tilt-induced Fermi surface deformation and the microscopic nature of impurity scattering, contrasting short-range defect potentials with long-range Coulomb interactions.

Our analysis reveals fundamentally different transport signatures depending on both the tilt regime and the impurity type. For short-range impurities, the conductivity increases as the system moves away from the critical point ($t=1$) in both the subcritical and overcritical regimes. Conversely, long-range impurities produce an enhanced conductivity near the critical point that decays as the tilt increases or decreases. Furthermore, while short-range scattering results in negligible or moderate energy dependence, long-range impurities lead to a strong energy dependence: nearly quadratic for $t < 1$ and linear for $t > 1$.

These differences originate from the distinct energy and tilt dependencies of the respective scattering rates. In the subcritical regime, the short-range scattering rate is proportional to the DOS, scaling linearly with energy and diverging as $t \to 1$. In contrast, the long-range scattering rate is nearly independent of tilt for $t < 1$ but exhibits a divergent energy dependence. In the overcritical regime, the short-range scattering rate scales linearly with $t$ with weak energy dependence, whereas the long-range scattering rate follows a $\sec^{-1}(\pm t)$ dependence while remains divergent with energy.

A defining feature of the Type II regime is the transition of the Fermi surface into open hyperbolas. To regularize the resulting divergent DOS in our low-energy effective model, we introduce a finite momentum cutoff $\Lambda$, a role naturally played by the Brillouin zone boundary in lattice models. For short-range impurities, this regime is characterized by extreme transport anisotropy ($\sigma_{yy} \gg \sigma_{xx}$), a universal hallmark of Type II Dirac fermions. For long-range Coulomb impurities, although the scattering rates are intrinsically regularized, the conductivity remains dependent on the momentum cutoff.

At the Lifshitz transition ($t=1$), macroscopic transport signatures diverge based on the microscopic potential. For short-range impurities, the van Hove singularity in the critical DOS drives a divergence in the scattering self-energy ($\sim 1/\sqrt{|\omega|}$), manifesting as a sharply localized conductivity dip. In contrast, for long-range scattering, the momentum-dependent potential regularizes this geometric divergence, resulting in a pronounced conductivity peak driven by the enhanced DOS. Crucially, we analytically demonstrate that vertex corrections vanish identically at this critical point for both impurity models. This result justifies the bare-bubble approximation as an exact description of lowest-order transport at the Lifshitz transition.

A comparison of our results with recent numerical localization studies \cite{NandaarXiv} reveals both illuminating convergences and instructive divergences. Both investigations underscore the pivotal role of the Lifshitz transition ($t=1$) as a critical point for transport. However, the physical origin and quantitative manifestation of these phenomena differ. In our work, the divergence of the scattering rate  in the bulk regime manifests as a sharp dip in the DC conductivity at the transition. In contrast, in the mesoscopic regime quantum interference leads to a pronounced spike in the scaling coefficient $\beta(g)$, signaling a critical point for localization.

Finally, we provided a profound physical intuition for these phenomena by reformulating the Hamiltonian tilt parameter as an effective metric curvature in momentum space. This geometric framework elegantly unifies our analytical results, demonstrating that the tilt-dependent scaling of the scattering lifetimes and conductivities arises naturally from the invariant volume element of the curved pseudo-Riemannian phase space.

\section{Summary}

This work provides a comprehensive study of DC conductivity in two-dimensional tilted Dirac semimetals across the subcritical, critical, and overcritical regimes. By contrasting short-range and long-range impurity scattering, we find very rich and fundamentally different behaviors. 
While for long-range impurities, in both
subcritical and overcritical regimes, the further we go from $t=1$ the conductance becomes suppressed, in the short-range case the transport is more profound away from critical point.
Another key difference shows in the energy dependence with no or weak energy dependence for short-range case in sub- and over-critical regimes, while we see quadratic and linear energy depedence, respectively, in the case of long-range impurities. Also, In the overcritical regime, we see a strong transport anisotropy and distinct energy-dependent scaling laws as universal hallmarks of Type II Dirac fermions, requiring a momentum cutoff to regularize the open Fermi surface. Finally, exactly at the Lifshitz transition ($t=1$) we see strong energy dependence which reveals a very localized conductivity dip around Dirac point for short-range defects. 

Beyond these signatures, we establish a unified geometric framework that reformulates the tilt parameter as an effective metric curvature in momentum space. This perspective reveals that the scaling of scattering lifetimes and conductivities emerges naturally from the invariant volume element of a curved phase space. Ultimately, our findings position the tilt parameter as a powerful tuning knob for driving topological phase transitions and engineering highly directional quantum transport, providing clear experimental benchmarks for future studies on tunable 2D Dirac and Type II semimetals.

\appendix

\section{Vertex corrections for short-range impurities}
\label{app:vertex}

The first-order vertex correction for the tilted Dirac Hamiltonian is given by:
\begin{align}\label{eq:vertex_renorm_shortrange}
\delta v_\alpha = n_{\rm imp} &\int \frac{d^2 q'}{(2\pi)^2} |V(\mathbf{q}-\mathbf{q}')|^2 \nonumber\\
& \qquad \times
G^R(\mathbf{q}',\omega) \, v_\alpha(\mathbf{q}') \, G^A(\mathbf{q}',\omega).
\end{align}
For short-range impurities, the potential is constant, $V(\mathbf{q}-\mathbf{q}') = V_0$. At a fixed positive energy $\omega > 0$, transport is dominated by the conduction band. Within the first Born approximation, the retarded and advanced Green's functions simplify to:
\begin{align}
G^{R/A}(\mathbf{q}, \omega) = \frac{(\omega - tq_y \pm i\Gamma)\sigma_0 - \tau_z(q_y\sigma_x + q_x\sigma_y)}{(\omega - tq_y \pm i\Gamma)^2 - (q_x^2+q_y^2)},
\end{align}
where $\Gamma \equiv \Gamma_{\text{II}}$ is simplified notation for the constant scattering rate for the conduction band.

Evaluating the product $G^R v_x G^A$ yields:
\begin{align}
\nonumber
    G^R v_x G^A &= A(\mathbf{q}, \omega) \, \tau_z\sigma_y + B(\mathbf{q}, \omega) \, \tau_0 \sigma_z\\
    & + C(\mathbf{q}, \omega) \, \tau_z \sigma_x + D(\mathbf{q}, \omega) \, \mathbb{I}_4,
\end{align}
with the coefficients defined as:
\begin{align}
\nonumber
A(\mathbf{q}, \omega) &=  \frac{(\omega - t q_y)^2+ q_x^2 - q_y^2+ \Gamma^2}{|D^R|^2}, \\
\nonumber
B(\mathbf{q}, \omega) &= \frac{2 \Gamma q_y}{|D^R|^2}, \\
\nonumber
C(\mathbf{q}, \omega) &=\frac{-2 q_x q_y}{|D^R|^2} , \\
D(\mathbf{q}, \omega) &= \frac{2 q_x (\omega - t q_y)}{|D^R|^2},
\end{align}
where 
\begin{align} |D^R|^2 = \big[ (\omega - t q_y)^2 - \Gamma^2 - q^2\big]^2 + 4\Gamma^2 (\omega - t q_y)^2. 
\end{align}
Since $|D^R|^2$ is even in $q_x$, the terms $C(\mathbf{q}, \omega)$ and $D(\mathbf{q}, \omega)$ vanish upon integration over $d^2q$. 

Furthermore, the term proportional to $B(\mathbf{q}, \omega)$ does not contribute to the longitudinal conductivity $\sigma_{xx}$ because its matrix structure leads to vanishing form factors from the eigenstates $  \bra{\lambda,{\bf q}} \tau_0 \sigma_z \ket{\lambda,{\bf q}}= 0$. Consequently, only the $A(\mathbf{q}, \omega)$ term survives. Because it shares the $\tau_z \sigma_y$ matrix structure of the bare velocity operator $v_x$, the vertex correction scalar-renormalizes the velocity operator $\mathcal{V}_x = v_x \big[1 + \mathcal{K} \big]$ with
\begin{align}\label{eq:K-shortrange}
\mathcal{K}  = \gamma_{\rm imp} \int \frac{d^2q}{2\pi} \, \frac{(\omega - t q_y)^2+ q_x^2 - q_y^2+ \Gamma^2}{|D^R|^2}.
\end{align}

\subsection{Analytical Evaluation in the Subcritical Regime ($t < 1$)}

To evaluate the integral for $\mathcal{K}$ analytically, we leverage the geometric transformation introduced in Sec.~\ref{sec:geometric}, mapping the tilted Dirac cone to an isotropic one. 
We define the new momentum variables
\begin{align}\label{eq:transform_q_to_p}
\begin{pmatrix}
    p_x \\ p_y \\ E
\end{pmatrix}
=
\begin{pmatrix}
1 &0&0\\
  0&  \sqrt{g} &  \frac{t}{\sqrt{g}} \\ 
    0&0 & 1
\end{pmatrix}
\begin{pmatrix}
    q_x \\ q_y \\ \omega
\end{pmatrix},
\end{align}
with metric determinant $g = 1 - t^2$,
Jacobian $d^2 q = \frac{1}{\sqrt{g}} dp_x dp_y$. 
The off-diagonal term in the transformation guarantees that
we get a isotropic quadratic form
\begin{align}
    (\omega - t q_y)^2 - (q_x^2 + q_y^2) =
    \frac{\omega^2}{g} -( p_x^2+ p_y^2)
\end{align}
Thus, the denominator in the integrand
of renormalization factor $\mathcal{K}$
becomes
\begin{align}
    |D^R|^2 &= (\omega^2/g - p^2 - \Gamma^2)^2 + 4\Gamma^2 E_t^2 \\
    \nonumber
    &\approx (\omega^2/g - p^2)^2 + 4\Gamma^2 E_t^2,
\end{align}
where $E_t=\omega-t q_y \equiv \omega/g-(t/\sqrt{g})p_y$. Assuming weak disorder limit ($\gamma_{\rm imp}\gg1$) we have neglected the small $\Gamma^2$ in comparison with $\omega^2/g$. 

Now, the integrand has a Lorentzian form which becomes strongly peaked at $p^2\sim p_\circ^2 =\omega^2/g$, allowing us to also simplify the numerator to
\begin{align}
    (\omega - t q_y)^2 + q_x^2 - q_y^2 + \Gamma^2 \approx 2q_x^2 = 2p^2 \cos^2\theta.
\end{align}
Substituting these into the integral for $\mathcal{K}(t)$ and separating the radial and angular components, we obtain:
\begin{align}
    \mathcal{K} \approx \frac{\gamma_{\rm imp}}{\sqrt{g}} & \int_0^{2\pi} \frac{d\theta}{ \pi }  \cos^2\theta  \nonumber \\
    & \times \int_0^\infty \frac{p^3 dp}{(p_\circ^2 - p^2)^2 + 4\Gamma^2 E_t^2(p,\theta)}.
\end{align}
The radial integral 
can be approximated as the sum over the singular resonant contribution at $p\sim p_\circ$ and the ultraviolet behavior at large $p$:
\begin{align}
    \approx  \frac{\pi p_\circ^2}{2 \Gamma E_t(p_\circ,\theta)} + {\cal O}\big[\log(\Lambda)\big].
\end{align}
Assuming very small scattering rate $\Gamma$ and the finite high-energy cutoff contribution can be ignored, yielding
\begin{align}
    \mathcal{K} \approx \frac{\gamma_{\rm imp} \omega}{ \Gamma \sqrt{g}} \int_0^{2\pi} \frac{d\theta}{ 2\pi }\frac{\cos^2\theta}{1 - t \sin\theta} .
\end{align}
Invoking the level broadening $\Gamma$ from Eq.~\eqref{eq.gamma1}, the vertex correction reduces to
\begin{align} \label{eq:K-undercritical}
    \mathcal{K}(t) = g  \int_0^{2\pi} \frac{d\theta}{2\pi}\frac{\cos^2\theta}{1 - t \sin\theta} 
= \frac{g}{(1 + \sqrt{g})},
\end{align}
which indicates that the velocity renormalization merely depends on defective spacetime geometry 
controlled by the tilt parameter $t$, independent of both dynamical parameters and impurity strength.

\subsection{Analytical Evaluation in the Overcritical Regime ($t > 1$)}

In the overcritical regime, the tilt parameter exceeds the effective Fermi velocity ($t > 1$), leading to a Lifshitz transition where the Fermi surface topology changes from a closed ellipse to an open hyperbola. To capture this geometry, we define the positive metric determinant $g = t^2 - 1 > 0$ and introduce the hyperbolic momentum transformation
\begin{align}
\begin{pmatrix}
    p_x \\ p_y \\ E
\end{pmatrix}
=
\begin{pmatrix}
1 &0&0\\
  0&  \sqrt{g} &  \frac{-t}{\sqrt{g}} \\ 
    0&0 & 1
\end{pmatrix}
\begin{pmatrix}
    q_x \\ q_y \\ \omega
\end{pmatrix},
\end{align}
In this pseudo-Riemannian geometry, the exact energy relation becomes $E_t^2 - (q_x^2 + q_y^2) = p_x^2 + p_y^2 - p_\circ^2$ with $p_\circ = \omega / \sqrt{g}$.

To evaluate the integrals over the open Fermi surface, we parameterize the momentum space using hyperbolic coordinates (rapidity $\eta$ and scaling factor $p$):
\begin{align}
    p_x = p \sinh\eta, \quad p_y = s \, p \cosh\eta,
\end{align}
where $s = \pm 1$ denotes the two disconnected branches of the hyperbola. The area element is $dp_x dp_y = p \, dp \, d\eta$. The unperturbed energy evaluates to:
\begin{align}
    E_t = -\frac{\omega}{g} - \frac{t}{\sqrt{g}} p_y = -\frac{\omega}{g} \big( 1 + s \, t \cosh\eta \big).
\end{align}
Following the same weak-disorder approximation used for $t < 1$, the integral is strongly peaked at the hyperbolic shell $p = p_\circ$. The numerator simplifies to $2q_x^2 \approx 2p_\circ^2 \sinh^2\eta$. Performing the radial-like integration over $p$ using the Lorentzian approximation yields $\pi / (2 \Gamma |E_t|)$. Summing over both branches $s = \pm 1$, the vertex correction integral reads
\begin{align}
    \mathcal{K}(t) &= \frac{\gamma_{\rm imp}}{2 \sqrt{g}} \sum_{s=\pm 1} \int_{-\eta_c}^{\eta_c} d\eta \, \frac{ p_\circ^2 \: \sinh^2\eta}{ \Gamma |E_t(\eta, s)|} \nonumber\\
    & = \frac{\gamma_{\rm imp} \: \omega}{2 \Gamma \sqrt{g}} \sum_{s=\pm 1} \int_{-\eta_c}^{\eta_c} d\eta \, \frac{\sinh^2\eta}{|1 + s \, t \cosh\eta|}.\label{eq:K-shortrange-t>1}
\end{align}

Now, unlike the subcritical regime where the angular integral was bounded ($0$ to $2\pi$), the rapidity $\eta$ extends to infinity for an open hyperbola. Therefore, the integral diverges and requires an ultraviolet momentum cutoff $\Lambda$. To evaluate the sum of the integrals over the two hyperbolic branches, we note that for $t > 1$ and $\cosh\eta \ge 1$, the terms inside the absolute values have definite signs: $1 + t \cosh\eta > 0$ and $1 - t \cosh\eta < 0$. Therefore, the absolute values can be resolved as $|1 \pm t \cosh\eta| = \pm (t \cosh\eta \pm 1)$. Summing the integrands for the two branches we obtain
\begin{align}
    \nonumber
    \sum_{s=\pm 1} \frac{\sinh^2\eta}{|1 + s \, t \cosh\eta|} 
    = \frac{2t \cosh\eta \sinh^2\eta}{t^2 \cosh^2\eta - 1}.
\end{align}
Since the resulting integrand is an even function of the rapidity $\eta$, the integral over the symmetric domain $[-\eta_c, \eta_c]$ simplifies to:
\begin{align}
    \mathcal{I} = 4t \int_{0}^{\eta_c} d\eta \, \frac{\cosh\eta \sinh^2\eta}{t^2 \cosh^2\eta - 1}.
\end{align}

We can now solve this exactly by introducing the variable substitution $x = \sinh\eta$, which implies $dx = \cosh\eta \, d\eta$ and $\cosh^2\eta = 1 + x^2$. The upper limit of integration transforms to $x_c = \sinh\eta_c$. Recalling that the metric determinant is $g = t^2 - 1$ when $t>1$, the integral is rewritten as a rational function:
\begin{align}
    \mathcal{I} = 4t \int_{0}^{x_c} dx \, \frac{x^2}{t^2(1 + x^2) - 1} = 4t \int_{0}^{x_c} dx \, \frac{x^2}{t^2 x^2 + g}.
\end{align}
This integral can be easily evaluated by adding and subtracting $g$ in the numerator:
\begin{align}
    \nonumber
    \mathcal{I} &=\frac{4}{t} \int_{0}^{x_c} dx \left( 1 - \frac{g}{t^2 x^2 + g} \right) \\
    &= \frac{4}{t} \left[ x_c - \frac{\sqrt{g}}{t} \arctan\left( \frac{t x_c}{\sqrt{g}} \right) \right].
\end{align}

To connect this mathematical result to the physical system, we must relate the rapidity boundary $x_c$ to the ultraviolet momentum cutoff $\Lambda$. Since the invariant momentum transforms as $p_x = p_\circ \sinh\eta$, which perfectly corresponds to the unmodified momentum $q_x$, setting the integration boundary to $q_x = \Lambda$ yields:
\begin{align}
    x_c = \sinh\eta_c = \frac{\Lambda}{p_\circ} = \frac{\Lambda \sqrt{g}}{\omega}.
\end{align}
Substituting this relation into the evaluated integral gives:
\begin{align}
    \mathcal{I} = \frac{4 \Lambda \sqrt{g}}{t \omega} - \frac{4\sqrt{g}}{t^2} \arctan\left( \frac{t \Lambda}{\omega} \right).
\end{align}

Finally, inserting $\mathcal{I}$ back into the expression for the vertex correction factor $\mathcal{K}(t)$, the terms elegantly simplify:
\begin{align}
    \mathcal{K}(t) 
    = 2 \frac{\gamma_{\rm imp} }{ \Gamma} \left[ \frac{\Lambda}{t} - \frac{\omega}{t^2} \arctan\left( \frac{t \Lambda}{\omega} \right) \right] \approx 2 \frac{\gamma_{\rm imp} }{ \Gamma}  \frac{\Lambda}{t} 
\end{align}
This expression elucidates the resolution of apparent cutoff dependence discussed in the main text. The
explicit linear dependence of 
vertex correction integral on the ultraviolet cutoff $\Lambda$, due to the open topology of the hyperbolic Fermi surface, cancels out with the same linear dependence of broadening function $\Gamma$ derived in Eq.~\eqref{eq:self_energies_overcritical}. Therefore, for large enough high-energy cutoff ($\Lambda \gg \omega$), we again obtain a purely geometric vertex correction 
\begin{align} \label{eq:K-overcritical}
    \mathcal{K}(t) =  \frac{2g}{t^2} \approx 
    2\frac{t^2-1}{t^2}  \equiv 2 \frac{g}{g+1}
\end{align}

\subsection{Vanishing vertex corrections at the Lifshitz Transition ($t \to 1$)}

The geometric formalism offers a direct proof for the vanishing of the vertex correction at the critical point $t = 1$. This can be explicitly seen from both expressions \eqref{eq:K-undercritical} and \eqref{eq:K-overcritical}, as $t \to 1$ and thus $g \to 0$. From a physical perspective, the effective Fermi surface stretches into two parallel lines at the Lifshitz Transition. This leads to a van Hove singularity in the density of states, causing the scattering rate $\Gamma$ to diverge. This divergent broadening function suppresses multiple-scattering contributions. As a result, at the Lifshitz transition, the vertex correction vanishes ($\mathcal{V}_x = v_x$), and the bare bubble approximation becomes exact to first order in the impurity concentration.

\section{Vertex Corrections for Long-Range Coulomb Potential}
\label{app:vertex_long_range}

The first-order vertex correction for a long-range Coulomb impurity follows the form of Eq. \eqref{eq:vertex_renorm_shortrange}, with 
$V(\mathbf{q}-\mathbf{q}') = \frac{e^2}{2\epsilon |\mathbf{q}-\mathbf{q}'|}$ explicitly coupling the incoming and outgoing momentum states. The resulting velocity correction is
\begin{align}
    \delta v_\alpha(\mathbf{q})
    &= n_{\rm imp} 
   \Big(\frac{e^2}{2\epsilon}\Big)^2 
    \int \frac{d^2 q'}{(2\pi)^2} \frac{G^R(\mathbf{q}', \omega) v_\alpha(\mathbf{q}') G^A(\mathbf{q}', \omega)}{|\mathbf{q}-\mathbf{q}'|^2}.
\end{align}
Following a procedure similar to that in Sec. \ref{app:vertex}, we find that
\begin{align}\label{eq:K-longrange}
\mathcal{K}({\bf q})  = 
n_{\rm imp} \alpha^2 \int d^2q' \, \frac{(\omega - t {q'}_y)^2+ {q'}_x^2 - {q'}_y^2+ \Gamma^2}{|D^R_{{\bf q}'}|^2\: |\mathbf{q}-\mathbf{q}'|^2},
\end{align}
which is similar to Eq. \eqref{eq:K-shortrange} but includes an additional term in the denominator arising from the long-range interaction kernel. 

Similar to the self-energy calculations in the main text, by focusing on the limit of small external momentum and employing a Lorentzian approximation for the integral, we obtain
\begin{align}
    \mathcal{K} \approx & \frac{n_{\rm imp} \alpha^2}{\sqrt{g}}  \int_0^{2\pi} d\theta  \cos^2\theta  \nonumber \\
    & \times \int_0^\infty \frac{2 p^3 dp}{(p_\circ^2 - p^2)^2 + 4\Gamma^2 E_t^2(p_\circ,\theta)}\frac{1}{E_t^2(p_\circ,\theta)},
\end{align}
for the subcritical case ($t<1$). In this regime, the energy terms in the transformed and original coordinates are related by $q^2=p^2-\omega^2/g+E_t^2$. At the Lorentzian pole ($p\sim p_\circ=\omega/\sqrt{g}$), this reduces to $q^2|_{p=p_\circ}=E_t^2(p_\circ,\theta)$, which leads to the additional factor of $E_t^2(p_\circ,\theta)$ in the denominator.
Neglecting the contribution from the high-energy cutoff as in Sec. \ref{app:vertex}, the Lorentzian approximation yields
\begin{align}
    \mathcal{K} & \approx  \frac{n_{\rm imp} \alpha^2}{\sqrt{g}}  \int_0^{2\pi} d\theta  \cos^2\theta  \: \frac{\pi p_\circ^2}{\Gamma\: E_t^3(p_\circ,\theta)} \nonumber\\
    &= 
    \frac{\pi n_{\rm imp} \alpha^2 g^{3/2}}{\Gamma\, \omega}  \int_0^{2\pi} d\theta  \cos^2\theta  \: \frac{1}{(1-t\sin\theta)^3}.
\end{align}
After substituting the broadening function, the prefactor outside the integral simplifies to $(2/\pi)g^{3/2}$. Since the integral evaluates to $\pi g^{-3/2}$, it follows that, within this limit and under our approximations, the vertex correction results in a constant enhancement of the velocity (${\cal K}={\rm cte}$) in the subcritical limit for long-range impurity.

Now, in the overcritical regime, 
we can readily see that the renormalization
integral looks like Eq. \eqref{eq:K-shortrange-t>1} with additional term $|E_{t}(\eta,s)|^2$ in the integrand corresponding to the $q^2$ evaluated at the Lorentzian pole. Therefore, we have 
\begin{align}
    \mathcal{K}(t) &= \frac{n_{\rm imp} \alpha^2}{2 \sqrt{g}} \sum_{s=\pm 1} \frac{1}{\Gamma_s} \int_{-\infty}^{\infty} d\eta \, \frac{ p_\circ^2 \: \sinh^2\eta}{  |E_t|^3(\eta, s)} \nonumber\\
    & = \frac{n_{\rm imp} \alpha^2}{2 |\omega|} \sum_{s=\pm 1} \frac{g^{3/2}}{\Gamma_s}\int_{-\infty}^{\infty} d\eta\, \frac{\sinh^2\eta}{ |1 + s \, t \cosh\eta|^3},\label{eq:K-longrange-t>1}
\end{align}
with $\Gamma_s$ given in Eq. 
\eqref{eq:Gamma-longrange-t>1} for the two bands.
The integrals above can be exactly evaluated as
\begin{align}
\frac{1}{(t^2-1)^{3/2}}
\left[\sec^{-1}(\pm t) \mp \frac{\sqrt{t^2-1}}{t^2} \right].
\end{align}
Then, by invoking the broadening functions from Eq. \eqref{eq:Gamma-longrange-t>1}
and substituting everything in Eq. \eqref{eq:K-longrange-t>1}, the renormalization due to the vertex corrections for overcritical regime of long-range impurity takes the explicit form:
\begin{align}
    {\cal K}_\pm \propto 1 \mp \frac{\sqrt{t^2-1}}{t^2 \sec^{-1} (\pm t) }
\end{align}
Crucially, unlike the subcritical regime, the vertex correction for $t>1$ is not a constant but exhibits a geometric dependence on the tilt parameter $t$. Most notably, taking the limit as the system approach the Lifshitz transition from the overcritical side ($t \rightarrow 1^+$), the asymptotic expansions of both $\sqrt{t^2-1}/t^2$ and $\sec^{-1} t$ converge to $\sqrt{2(t-1)}$. Consequently, the second term approaches unity, resulting in an exact cancellation of the vertex correction at Lifshitz transition. As $t \to \infty$, the correction monotonically approaches a constant value, confirming that the tilt dynamically tunes the impact of multiple-scattering events across the Type II phase.




\section{Scaling Analysis of Transport Anisotropy in the Overcritical Regime for short-range potential}
\label{app:anisotropy}

In this appendix, we derive the transport scaling for $t>1$. In particular, we show why $\sigma_{xx}$ is non-monotonic, reaching a maximum at $t=\sqrt{2}$ and decaying as $1/t$ for large $t$, whereas $\sigma_{yy}$ grows monotonically.
For $\Lambda \gg \omega$, the Fermi energy is negligible compared with the kinetic and tilt terms, so the on-shell condition becomes
\begin{align}
    \varepsilon_{\pm}(\mathbf q) = t q_y \pm \sqrt{q_x^2+q_y^2} \approx 0.
\end{align}
In this regime, the dominant contribution comes from momenta near the cutoff, where the open Fermi surface approaches its asymptotic straight lines. 
Therefore, we may safely neglect the finite small energy and use the asymptotic line equations 
$q_x = \pm \sqrt{t^2-1}  q_y$
from the on-shell condition to determine both the velocities and the integration limits.
Particularly, the corresponding group velocities read
\begin{align}
    v_{y,\pm} &= \frac{\partial \varepsilon}{\partial q_y}
    = t\pm\frac{q_y}{\sqrt{q_x^2+q_y^2}}
    = \frac{t^2-1}{t},\\
    v_{x,\pm} &= \frac{\partial \varepsilon}{\partial q_x}
    = \pm \frac{q_x}{\sqrt{q_x^2+q_y^2}}
    = \mp \frac{\sqrt{t^2-1}}{t}.
\end{align}

Now within the Kubo formula, the DC conductivity is proportional to the integral of $v_i^2$ over the Fermi surface. Integrating over $q_y$ using $\delta(\varepsilon)$ gives the Jacobian
\begin{align}
    \int d^2q\,\delta(\varepsilon)\,\cdots
    = \int dq_x\,\frac{1}{|v_y|}\,\cdots
    = \int dq_x\,\frac{t}{t^2-1}\,\cdots .
\end{align}

The radial cutoff $q_x^2+q_y^2\le \Lambda^2$ determines the effective range of $q_x$. Using the on-shell condition $q_y^2=q_x^2/(t^2-1)$, we obtain
\begin{align}
    q_x^2\left(1+\frac{1}{t^2-1}\right)\le \Lambda^2,
\end{align}
which implies that
 the effective cutoff 
 for integration over momentum $q_x$ is $\Lambda_x=\Lambda \sqrt{t^2-1}/t$.
It should be noted that the order of integration over $q_x$ and $q_y$ can be interchanged
which results in $ \int d^2q\,\delta(\varepsilon)\,\cdots
= \int dq_y\,\frac{1}{|v_x|}\,\cdots$. But once can easily verifies that since $|v_x|\Lambda_x = |v_y|\Lambda_y$, 
we get the identical results by swapping the order of integration over two momenta.

We can now extract the conductivity scaling. For $\sigma_{xx}$,
\begin{align}
    \sigma_{xx}
    &\propto \int_{-\Lambda_x}^{\Lambda_x} dq_x\,
    \frac{t}{t^2-1} \, v_x^2 = 2 \frac{\Lambda_x}{t}
    = 2\Lambda\frac{\sqrt{t^2-1}}{t^2}.
\end{align}
By differentiating with respect to $t$ one can easily finds the maximum value of conductivity is attained for $t=\sqrt{2}$.
For large tilting $t\gg1$ the conductivity asymptotically vanishes as $\sigma_{xx}\sim 1/t$. Both these features are clearly seen in the numerical results shown in Fig. \ref{fig:anisotropy}.   
For $\sigma_{yy}$, the same analysis gives
\begin{align}
    \sigma_{yy}
    &\propto \int_{-\Lambda_x}^{\Lambda_x} dq_x\,
    \frac{t}{t^2-1} \, v_y^2 =  2\Lambda\frac{(t^2-1)^{3/2}}{t^2}.
\end{align}
Its derivative is positive for all $t>1$, so $\sigma_{yy}$ increases monotonically.
Therefore, in the overcritical regime, $\sigma_{xx}$ is suppressed at large tilt while $\sigma_{yy}$ is enhanced, providing a simple asymptotic explanation for the strong transport anisotropy of the Type II Dirac phase.

\bibliography{refs.bib}
\end{document}